\documentclass{article} 
\usepackage[preprint]{colm2026_conference}

\usepackage{microtype}
\usepackage{hyperref}
\usepackage{url}
\usepackage{booktabs}

\usepackage{lineno}
\usepackage[table]{xcolor}
\usepackage{amsmath}
\usepackage{multirow}

\definecolor{darkblue}{rgb}{0, 0, 0.5}
\hypersetup{colorlinks=true, citecolor=darkblue, linkcolor=darkblue, urlcolor=darkblue}
\usepackage{caption}
\usepackage[T1]{fontenc}
\usepackage{tcolorbox}
\usepackage{listings}
\usepackage{enumitem} 
\tcbuselibrary{listings, skins, breakable}

\definecolor{preEditColor}{RGB}{220,220,220}
\definecolor{s2ErrorColor}{RGB}{173,216,230}
\definecolor{severeColor}{RGB}{255,182,182}
\definecolor{attrErrorColor}{RGB}{255,200,150}
\definecolor{assertErrorColor}{RGB}{255,235,150}
\definecolor{workaroundColor}{RGB}{180, 120, 0}

\usepackage{upquote}
\lstdefinestyle{codestyle}{
    basicstyle=\ttfamily\footnotesize,
    breaklines=false,
    showstringspaces=false,
    language=Python,
    escapeinside={(*@}{@*)},
    upquote=true,
}
\newcommand{\oldapi}[1]{\textcolor{red}{\textbf{#1}}}

\usepackage{fancyvrb}

\newcommand{\targetapi}[1]{\textcolor{blue}{\textbf{#1}}}

\title{Understanding Robustness of Model Editing in Code LLMs}

\author{%
  Vinaik Chhetri$^1$ \quad
  Moghis Fereidouni$^2$ \quad
  A.B. Siddique$^2$ \quad
  Umar Farooq$^1$ \\[0.4em]
  $^1$Louisiana State University, Louisiana \quad
  $^2$University of Kentucky, Kentucky \\[0.2em]
  \texttt{\{vchhet2, ufarooq\}@lsu.edu} \quad
  \texttt{\{moghis.fereidouni, siddique\}@cs.uky.edu}
}

\newcommand{\stitle}[1]
{\noindent\textup{\textbf{#1}}}

\begin{document}

\ifcolmsubmission
\linenumbers
\fi

\maketitle

\begin{abstract}

Large language models~(LLMs) for code are increasingly used in software development, but they remain static after pretraining while APIs and software libraries continue to evolve. 
Model editing offers a lightweight alternative to retraining for incorporating API updates, yet it remains unclear whether existing editing methods can induce correct API migration, generalize that behavior to unseen tasks, and preserve performance on tasks involving unmodified APIs. 
We present a controlled benchmark for evaluating model editing under API updates in code LLMs, built from HumanEval, MBPP, and APPS, with 2,040 problems spanning 140 unique synthetic API modifications, together with an execution sandbox that enforces edited APIs under standard Python semantics. 
We evaluate several state-of-the-art editing methods on three code LLMs under both single-edit and successive-edit regimes using execution-based metrics that distinguish successful API adoption from workaround-based task completion.
Under single edits, edited models generalize poorly to unseen uses of the modified API, and many apparent successes are workaround-based rather than true API migrations. 
Performance on tasks involving unmodified APIs also degrades, although memory-based methods and fine-tuning preserve specificity better than locate-then-edit methods. 
Under successive edits, most method-model combinations collapse to near-zero Pass@k on both generalization and specificity, revealing substantial interference beyond the target edits. 
A two-factor Shapley decomposition further shows that single-edit failures on generalization include a substantial compilation component, whereas specificity failures are more often post-compilation. 
Under successive edits, failures become predominantly compilation-driven. 
Finally, we develop an error taxonomy that separates successful adoption from workaround-based success and localizes failures to compilation, API adaptation, execution, and behavioral stages.
\end{abstract}

\section{Introduction}
\label{sec:intro}

Large language models~(LLMs) for code are increasingly used in software development workflows such as code completion, bug fixing, and program transformation~\citep{github-2021-copilot, amazon-2023-codewhisperer}. 
However, these models remain static after pretraining, while the APIs and software libraries continue to evolve. 
As shown in Table~\ref{tab:deprecation-stats}, software ecosystems such as Android, pandas, NumPy, and Node.js introduce substantial numbers of API updates over relatively short release intervals.
Retraining can, in principle, absorb such changes, but it is often expensive and operationally cumbersome. 
Model editing offers a lightweight alternative by updating targeted behavior after deployment~\citep{hartvigsen2023aging,meng2022locating}. 
Yet it remains unclear whether existing editing methods can reliably adapt code LLMs to evolving APIs without degrading their broader code generation capabilities.

Editing code LLMs poses a distinct challenge from standard knowledge editing. 
In code generation, correctness is determined by execution rather than surface-form similarity, and failures may arise from either syntactically invalid code or semantically incorrect programs.
Programming tasks also admit multiple functionally valid solutions. 
As a result, task success alone does not imply successful editing, since a model may satisfy the tests while bypassing the updated API through an alternative implementation.

Existing work does not adequately address this challenge.
Some studies evaluate real API changes across library versions, where uncontrolled pretraining exposure makes it difficult to attribute the model's post-edit performance to successful editing rather than prior memorization~\citep{lin2025lightweightmodeleditingllms,wang2025codesyncsynchronizinglargelanguage,wang2025llmsmeetlibraryevolution}. 
Others rely on surface-form metrics such as Exact Match, BLEU, or ROUGE~\citep{10.1109/ICSE55347.2025.00049,lin2025lightweightmodeleditingllms,gu2024neuronpatchingsemanticbasedneuronlevel,wang2025llmsmeetlibraryevolution}, or on task success alone~\citep{liu2025codeupdatearenabenchmarkingknowledgeediting,wang2025codesyncsynchronizinglargelanguage,liu2025cremerobustnessenhancementcode}.
Prior evaluations also omit settings in which successive edits accumulate over time.
As a result, several key questions remain unanswered: whether edited models can reliably generate syntactically and semantically correct code, whether they generalize the updates to unseen problems, whether they preserve specificity, and whether they remain stable as updates span multiple APIs over time.
Addressing these questions requires a controlled evaluation setting that isolates editing effects, minimizes confounding from pretraining exposure, and makes API adaptation measurable.

\begin{table}[t]
\centering
\caption{Number of API updates in major software ecosystems between~releases.}
\label{tab:deprecation-stats}
\vspace{-10pt}
\footnotesize
\begin{tabular}{llcc}
\toprule
\textbf{Ecosystem} & \textbf{Type} & \textbf{Versions} & \textbf{\# Changes} \\
\midrule
Android~\citep{android-api-diff-35} & Mobile framework & 34$\rightarrow$35 & 834 \\
NumPy~\citep{numpy-2.0.0-notes} & Scientific library & 1.9$\rightarrow$2.0 & 104 \\
Node.js~\citep{nodejs-v20-deprecations} & JavaScript runtime & 23$\rightarrow$24 & 196 \\
pandas~\citep{pandas-1.5.0-deprecations}& Data analysis library & 1.0$\rightarrow$1.5 & 200 \\
\bottomrule
\end{tabular}
\vspace{-22pt}
\end{table}

To address these gaps, we construct a controlled benchmark for evaluating API updates in code LLMs using synthetic API modifications derived from HumanEval~\citep{chen2021evaluating}, MBPP~\citep{google2025mbpp}, and APPS~\citep{hendrycksapps2021}, comprising 2,040 problems spanning 140 unique synthetic API modifications.
These synthetic updates allow us to isolate editing effects by minimizing confounds arising from pretraining exposure to real API changes.
We pair this benchmark with an execution sandbox that replaces original APIs with their updated versions while preserving standard Python semantics, ensuring that generated code is evaluated against the modified APIs.
We evaluate seven editing methods, Constraint Fine-Tuning (FT-L)~\citep{ft}, GRACE~\citep{hartvigsen2023aging}, A-GRACE~\citep{10.1109/ICSE55347.2025.00049}, MALMEN~\citep{tan2024massiveeditinglargelanguage}, MEMIT~\citep{meng2023memit}, PMET~\citep{li2024pmet}, and ROME~\citep{meng2022locating}, across three open-source code LLMs: CodeLlama-7B~\citep{codellama}, CodeQwen1.5-7B~\citep{qwen}, and DeepSeek-Coder-6.7B~\citep{deepseek-coder}. 
We study two editing regimes, \emph{single edit}, where one API change is introduced in isolation, and \emph{successive edits}, where edits accumulate over time.

Our empirical results show significant robustness gaps in current editing methods. 
In the single-edit setting, generalization performance drops substantially on unseen problems requiring the modified API, with memory-based methods and fine-tuning generally degrading less than locate-then-edit approaches, especially on the specificity set of tasks involving unmodified APIs. 
Importantly, many apparent successes are not genuine API migrations, since models often pass the tests through workaround-based solutions that avoid the edited interface. 
A two-factor Shapley decomposition~\citep{lundberg2017unified} further shows that failure mode differs between the two evaluation subsets: errors on generalization have a larger compilation component, whereas errors on specificity arise more often after compilation. 
Under successive edits, these failure modes become substantially more severe, and most method-model combinations become unstable on both generalization and specificity, often collapsing to near-zero Pass@k. 
To make these failure modes explicit, we develop an error taxonomy that separates successful adoption from workaround-based success and localizes failures to distinct stages of the execution pipeline.

This work makes the following contributions:
\begin{itemize}[leftmargin=*, itemsep=0pt, topsep=-2pt]
    \item We introduce a controlled benchmark for evaluating API updates in code LLMs together with an execution sandbox that enforces updated APIs under Python semantics.
    \item We present a systematic empirical study across multiple editing methods, code LLMs, and editing regimes, showing that current methods do not enable robust API updates.
    \item We develop a taxonomy of post-edit outcomes for generated code, localizing failures to compilation, API adaptation, execution, and behavioral stages.
\end{itemize}

\section{Related Work}

Automating API migration has been studied through example-based update tools~\citep{Lamothe_2022, 10.1145/3293882.3330571, 9054860, haryono2020automaticandroiddeprecatedapiusage, haryono2021androevolveautomatedupdateandroid}, signature-difference inference~\citep{9609153}, and more recently through LLM-based prompting~\citep{mahmud2024automatedupdateandroiddeprecated}. 
These approaches aim to translate code from deprecated APIs to their replacements, but they assume that the replacement API is available at inference time~\citep{wang2025llmsmeetlibraryevolution}. 
Closely related work has also examined deprecated API usage in code LLMs and shown that stale training data can cause models to continue producing obsolete interfaces~\citep{wang2025llmsmeetlibraryevolution}, motivating methods that update model behavior after deployment rather than relying solely on prompting.

The closest line of work studies model editing for code~\citep{lin2025lightweightmodeleditingllms}. 
However, existing evaluations differ from ours in several important ways. 
Some use real API changes, making it difficult to separate the effect of editing from possible pretraining exposure~\citep{wang2025codesyncsynchronizinglargelanguage, wang2025llmsmeetlibraryevolution}. 
Others rely on surface-form metrics~\citep{10.1109/ICSE55347.2025.00049, lin2025lightweightmodeleditingllms, gu2024neuronpatchingsemanticbasedneuronlevel}, or on task success alone~\citep{liu2025codeupdatearenabenchmarkingknowledgeediting, wang2025codesyncsynchronizinglargelanguage, liu2025cremerobustnessenhancementcode}, which cannot distinguish true API adoption from workaround-based success or localize failures across compilation and execution stages.
In addition, successive-edit settings remain largely unexplored for code LLMs.
In contrast, our benchmark uses synthetic API modifications to reduce pretraining confounds and evaluates a broader range of editing methods with execution-based measures that distinguish true API adoption from workaround behavior and decompose failures into compilation and post-compilation stages.

\section{Materials and Methods}

\subsection{Dataset}
We build our benchmark from three Python coding datasets, APPS, HumanEval, and MBPP, as summarized in Table~\ref{tab:dataset-stats}. From originally provided canonical solutions, we extract API usage patterns spanning 213 unique APIs, which yield 21,430 problem--API pairs. We then apply stratified proportional sampling to select 2,040 instances, producing a final benchmark with 1,875 problems and 140 unique synthetic APIs (more details in Appendix~\ref{app:dataset}).

\stitle{Split Design.} 
We divide the benchmark into three splits with distinct roles. 
For each instance, we identify a target API in the canonical solution and apply the corresponding modification to obtain a modified canonical solution, which serves as the edit target.
(1)~\emph{Reliability} provides the editing or fine-tuning instances and is used only to apply the edit, 
(2)~\emph{Generalization} evaluates whether the edited model correctly employs the updated API on unseen problems, and 
(3)~\emph{Specificity} contains problems involving unmodified APIs and evaluates whether editing preserves performance outside the target behavior.
Reliability, Generalization, and Specificity contain 469, 647, and 924 instances, respectively.

\stitle{API Modifications.} We apply five synthetic modification types: R1 (renaming), S1--S4 (parameter additions, reordering, and signature changes), detailed in Table~\ref{tab:api-mods}. Each modification transforms the target API call in the canonical solution into its new variant by updating the function signature, adding a required argument, or reordering parameters, producing the modified canonical solution used as the editing target.

\begin{table*}[t]
\centering
\footnotesize
\caption{Synthetic API modification types used in our benchmark. Bold text highlights the modified portions of the API interface.}
\vspace{-0.5em}
\begin{tabular}{
>{\raggedright\arraybackslash}p{0.4cm}
>{\raggedright\arraybackslash}p{3.4cm}
>{\raggedright\arraybackslash}p{3.2cm}
>{\raggedright\arraybackslash}p{4.3cm}
}
\toprule
\textbf{Type} & \textbf{Change} & \textbf{Original API} & \textbf{Modified API} \\
\midrule

R1 & API Rename
& \texttt{tuple(iterable)}
& \texttt{\textbf{create\_tuple}(iterable)}
 \\

S1 & Add optional parameter
& \texttt{all(iterable)}
& \texttt{all(iterable, \textbf{strict=False})}
\\

S2 & Add required parameter
& \texttt{heappush(heap,item)}
& \texttt{heappush(heap,item,\textbf{log})}
 \\

S3 & Reorder parameters
& \texttt{isinstance(x,int)}
& \texttt{isinstance(\textbf{int,x})}
\\

S4 & Change return type
& \texttt{max(nums) -> int}
& \texttt{max(nums) -> \textbf{(value,index)}}
\\

\bottomrule
\end{tabular}
\vspace{-1em}
\label{tab:api-mods}
\end{table*}

\subsection{Editing Methods and Models}
We evaluate seven editing methods across four paradigms: (1) \emph{Fine-tuning}: FT-L directly updates a targeted subset of parameters with norm constraints to limit interference with unrelated knowledge; (2) \emph{Memory-based}: GRACE and A-GRACE maintain a frozen base model with an external codebook, avoiding weight modification, with A-GRACE adding a contrastive MLP encoder for improved generalization; (3) \emph{Locate-then-Edit}: ROME, MEMIT, and PMET use causal tracing to identify and directly update FFN weight matrices, with MEMIT extending to batch edits across multiple layers and PMET isolating FFN hidden states to avoid contamination from attention representations; (4) \emph{Meta-learning}: MALMEN trains a hyper-network to generate parameter shifts, formulating multi-edit aggregation as a least squares problem to resolve conflicting updates across simultaneous edits while decoupling hyper-network computation from the base LM to reduce memory overhead. 

\stitle{Models.}
We use three open-source code LLMs representing diverse training strategies: \textit{CodeLlama-7B-Instruct}, an instruction-tuned variant trained on code; \textit{CodeQwen1.5-7B-Chat}, combining general-purpose pretraining with code instruction tuning; and \textit{DeepSeek-Coder-6.7B-Instruct}, trained on large-scale software repositories with instruction tuning.

\subsection{Code Execution Sandbox}
\label{sec:sandbox}
To evaluate whether models adapt correctly to modified APIs, we build a code-execution sandbox that enforces the edited interfaces while preserving standard Python semantics. This ensures that evaluation targets API adaptation rather than changes to the language itself. The sandbox simulates API evolution by replacing selected Python APIs with edited versions defined by our benchmark. These edits cover the modification families in Table~\ref{tab:api-mods}, while all other parts of the Python environment remain unchanged.

For each benchmark instance, we replace the relevant original APIs with their edited counterparts and execute model-generated code in a fresh Python interpreter. Because API resolution occurs against the edited definitions, any API call in the generated code is checked against the updated interface rather than the original one.

We evaluate each generated code instance in two stages. 
First, we apply a compilation-stage check that combines Python's built-in \texttt{compile()} function with verification of signature and interface conformance in the edited environment.
If this stage succeeds, we execute the code with unit tests derived from the original datasets to verify behavior under the updated APIs. This setup separates failures due to syntactic or interface-level incompatibility from downstream execution and behavioral failures while measuring functional correctness under updated interfaces.

\begin{table}[t]
\centering
\caption{API usage statistics across MBPP, HumanEval, and APPS, and their combination to use in our benchmark.}
\label{tab:dataset-stats}
\vspace{-0.5em}
\footnotesize
\setlength{\tabcolsep}{3pt}
\begin{tabular}{lcccc}
\toprule
\textbf{Metric} & \textbf{MBPP} & \textbf{HE} & \textbf{APPS} & \textbf{Combined} \\
\midrule
Number of problems & 191 & 26 & 1658 & 1,875\\
Problem–API pairs  & 200 & 26 & 1814 & 2,040\\
Unique synthetic APIs & 57& 16 & 128 & 140\\
\bottomrule
\end{tabular}
\vspace{-1em}
\end{table}

\section{Experimental Environment}

\stitle{Editing Regimes.}
Editing methods should be evaluated not only on isolated single updates but also on how they behave when multiple edits must coexist in the same model. 
Success on a single API change does not guarantee robustness under multiple updates, where interference, saturation, or instability may emerge only after edits accumulate. 
We therefore evaluate all methods under two complementary regimes.

\begin{itemize}[leftmargin=*, itemsep=0pt, topsep=2pt]
    \item \textbf{Single edit} applies one API modification in isolation and tests whether a method can incorporate that change without disrupting unrelated behavior.
    \item \textbf{Successive edits} applies multiple API updates cumulatively, testing whether methods remain stable as edits accumulate.
\end{itemize}

\stitle{Method Configurations.} We provide editing methods' configurations in Appendix~\ref{app:methods_config}.

\stitle{Evaluation Metrics.}
We generate $n=5$ samples per problem and report four metrics at $k \in {1,3,5}$. \emph{(1)~Compiles@$k$} measures the probability that at least one of the $k$ sampled outputs compiles. \emph{(2)~Pass@$k$} measures the probability that at least one sampled output passes all test cases.  
We further partition Pass@k into two disjoint problem-level outcomes. \emph{(3)~Successful Adoption@k} counts problems for which at least one of the k samples passes all tests while correctly using the edited API. \emph{(4)~Workaround@k} counts problems for which at least one sample passes all tests, but none of the passing samples correctly use the edited API.
This decomposition lets us distinguish apparent success through workaround behavior from genuine adoption of the updated API.

\stitle{Shapley Attribution.} To identify whether an editing method fails by 
producing uncompilable code or syntactically valid but incorrect code, we decompose $\texttt{Pass@k} = C \cdot Q$, where $C = \texttt{Compiles@k}$ is 
compilability and $Q = P(\text{Pass} \mid \text{Compiled})$ is post-compilation 
correctness. 
We compare an edited model (subscript $1$) to the non-edit baseline 
(subscript $0$), where the total change $\Delta P = P_1 - P_0$ reflects 
interacting contributions from both $C$ and $Q$, making it ambiguous to attribute 
$\Delta P$ to either factor alone. 
We apply two-factor Shapley attribution 
to disentangle these contributions, yielding $\phi_C$ and $\phi_Q$ such that 
$\Delta P = \phi_C + \phi_Q$. 
We attribute the performance change, $\Delta P$, to the compilation failure contribution via \textit{CF}~$= |\phi_C| / (|\phi_C| + |\phi_Q|)\times 100$, where values near $100$ indicate performance changes are driven by compilation failures and values near $0$ indicate post-compilation failures. 
See complete derivation in Appendix~\ref{app:shapley_der}.

\section{Results}
\subsection{Single Edits}
\label{sec:single-edit}

\begin{table}[t]
\centering
\scriptsize
\caption{Model performance for single edits across Generalization and Specificity. Aggregate $\Delta P$ and CF are averaged across models.}
\vspace{-0.5em}
\setlength{\tabcolsep}{2.5pt}
\begin{tabular}{l|cc|cc|cc|cc|cc|cc|cc|cc}
\toprule
& \multicolumn{8}{c|}{\textbf{Generalization}} & \multicolumn{8}{c}{\textbf{Specificity}} \\
\cmidrule(lr){2-9} \cmidrule(lr){10-17}
& \multicolumn{2}{c|}{\textbf{CodeLlama}} & \multicolumn{2}{c|}{\textbf{CodeQwen}} & \multicolumn{2}{c|}{\textbf{DeepSeek}} & \multicolumn{2}{c|}{\textbf{Average}} & \multicolumn{2}{c|}{\textbf{CodeLlama}} & \multicolumn{2}{c|}{\textbf{CodeQwen}} & \multicolumn{2}{c|}{\textbf{DeepSeek}} & \multicolumn{2}{c}{\textbf{Average}} \\
\cmidrule(lr){2-3} \cmidrule(lr){4-5} \cmidrule(lr){6-7} \cmidrule(lr){8-9}
\cmidrule(lr){10-11} \cmidrule(lr){12-13} \cmidrule(lr){14-15} \cmidrule(lr){16-17}
\textbf{Method} & C@5 & P@5 & C@5 & P@5 & C@5 & P@5 & $\Delta$P & CF & C@5 & P@5 & C@5 & P@5 & C@5 & P@5 & $\Delta$P & CF \\
\midrule
pre-edit & 1.00 & 1.00 & 1.00 & 1.00 & 1.00 & 1.00 & -- & -- & 1.00 & 1.00 & 1.00 & 1.00 & 1.00 & 1.00 & -- & -- \\
\midrule
FT-L & 0.76 & 0.63 & 0.76 & 0.68 & 0.82 & 0.66 & -0.34 & 59.97 & 1.00 & 0.88 & 1.00 & 0.92 & 0.96 & 0.80 & -0.13 & 5.72 \\
GRACE & 0.74 & 0.64 & 0.77 & 0.69 & 0.83 & 0.66 & -0.34 & 59.93 & 1.00 & 0.92 & 1.00 & 0.93 & 0.95 & 0.83 & -0.11 & 9.26 \\
A-GRACE & 0.75 & 0.62 & 0.76 & 0.66 & 0.81 & 0.64 & -0.36 & 57.97 & 1.00 & 0.90 & 1.00 & 0.94 & 0.94 & 0.82 & -0.12 & 12.48 \\
MALMEN & 0.71 & 0.44 & 0.73 & 0.51 & 0.79 & 0.57 & -0.49 & 43.08 & 0.95 & 0.68 & 0.91 & 0.67 & 0.89 & 0.61 & -0.34 & 20.00 \\
MEMIT & 0.50 & 0.26 & 0.73 & 0.58 & 0.45 & 0.23 & -0.64 & 54.60 & 0.66 & 0.37 & 0.95 & 0.81 & 0.51 & 0.28 & -0.51 & 40.21 \\
PMET & 0.62 & 0.32 & 0.74 & 0.65 & 0.49 & 0.25 & -0.59 & 54.29 & 0.79 & 0.47 & 0.98 & 0.91 & 0.60 & 0.37 & -0.42 & 34.57 \\
ROME & 0.66 & 0.35 & 0.74 & 0.55 & 0.42 & 0.26 & -0.61 & 51.27 & 0.87 & 0.45 & 0.96 & 0.75 & 0.57 & 0.36 & -0.48 & 29.63 \\
\bottomrule
\end{tabular}
\vspace{-1em}
\label{tab:instant-results-gen-spec-at5}
\end{table}

Single edits degrade code generation performance through both compilation and post-compilation failures. On the Generalization set (Table~\ref{tab:instant-results-gen-spec-at5}), both Compiles@5 and Pass@5 decline after editing. Compiles@5 drops from a baseline of 1.00 to 0.42--0.83, while Pass@5 drops more sharply to 0.23--0.69. 
Memory-based methods and fine-tuning remain the most resilient, with GRACE, A-GRACE, and FT-L achieving Compiles@5 of 0.74--0.83 and Pass@5 of 0.62--0.69 across models. In contrast, locate-then-edit methods show larger degradation, with Compiles@5 falling as low as 0.42--0.74 and Pass@5 as low as 0.23--0.65. MALMEN lies between these extremes, reaching 0.71--0.79 in Compiles@5 and 0.44--0.57 in Pass@5. The effect also varies by model. CodeQwen remains the most robust under editing, while CodeLlama and DeepSeek degrade much more severely under locate-then-edit approaches.

On the Specificity set, performance also declines, indicating regression on unrelated tasks. Compiles@5 remains relatively high for the strongest methods, ranging from 0.94--1.00 for FT-L, GRACE, and A-GRACE, but drops substantially for locate-then-edit methods, reaching as low as 0.51--0.98. 
Pass@5 shows larger declines, falling from 1.00 to 0.28--0.94. Overall, memory-based methods and fine-tuning preserve performance best, with Pass@5 of 0.80--0.94, whereas locate-then-edit methods fall to 0.28--0.91. DeepSeek is the most affected model in this setting, dropping to 0.28--0.37 under locate-then-edit methods. MALMEN remains intermediate, with Compiles@5 of 0.89--0.95 and Pass@5 of 0.61--0.68. 
To summarize, these results show that even a single edit can both reduce the likelihood of generating compilable code and degrade final task performance, including on unrelated tasks.

To understand the source of these declines, we decompose the Pass@5 drop into compilation and post-compilation contributions using Shapley attribution.
Table~\ref{tab:instant-results-gen-spec-at5} also reports CF, the share of overall degradation attributable to compilation. 
On the Generalization subset, compilation is the dominant failure source for every method, with CF ranging from 43.08\% (MALMEN) to 59.97\% (FT-L). 
On the Specificity subset, these values are substantially lower, ranging from 5.72\% (FT-L) to 40.21\% (MEMIT), indicating that degradation on unrelated tasks more often arises after successful compilation. 
This contrast reveals two failure regimes. Generalization failures are driven primarily by reduced compilability after editing, whereas Specificity failures more often reflect downstream runtime or behavioral errors after compilation.

\subsection{Workarounds and True API Adoption}
\label{sec:workarounds}

\begin{table}[t]
\centering

\scriptsize
\caption{Model performance for single edits on the canonically filtered Generalization set.}
\vspace{-0.5em}
\setlength{\tabcolsep}{5pt}
\begin{tabular}{l|cccc|cccc|cccc}
\toprule
& \multicolumn{4}{c|}{\textbf{CodeLlama}} & \multicolumn{4}{c|}{\textbf{CodeQwen}} & \multicolumn{4}{c}{\textbf{DeepSeek}} \\
\cmidrule(lr){2-5} \cmidrule(lr){6-9} \cmidrule(lr){10-13}
\textbf{Method} & C@5 & P@5 & A@5 & W@5 & C@5 & P@5 & A@5 & W@5 & C@5 & P@5 & A@5 & W@5 \\
\midrule
pre-edit & 1.00 & 1.00 & -- & -- & 1.00 & 1.00 & -- & -- & 1.00 & 1.00 & -- & -- \\
\midrule
FT-L & 0.51 & 0.40 & 0.17 & 0.23 & 0.57 & 0.46 & 0.18 & 0.29 & 0.60 & 0.38 & 0.17 & 0.21 \\
GRACE & 0.49 & 0.41 & 0.13 & 0.28 & 0.58 & 0.48 & 0.17 & 0.31 & 0.64 & 0.38 & 0.19 & 0.19 \\
A-GRACE & 0.50 & 0.43 & 0.17 & 0.26 & 0.57 & 0.44 & 0.16 & 0.27 & 0.60 & 0.34 & 0.15 & 0.19 \\
MALMEN & 0.51 & 0.28 & 0.11 & 0.17 & 0.59 & 0.35 & 0.10 & 0.25 & 0.60 & 0.36 & 0.17 & 0.19 \\
MEMIT & 0.34 & 0.16 & 0.10 & 0.06 & 0.58 & 0.38 & 0.15 & 0.23 & 0.43 & 0.19 & 0.11 & 0.08 \\
PMET & 0.49 & 0.20 & 0.10 & 0.10 & 0.55 & 0.44 & 0.16 & 0.29 & 0.45 & 0.17 & 0.13 & 0.04 \\
ROME & 0.54 & 0.26 & 0.09 & 0.17 & 0.62 & 0.37 & 0.13 & 0.24 & 0.42 & 0.19 & 0.11 & 0.08 \\
\bottomrule
\end{tabular}
\vspace{-1em}
\label{tab:canonical-instant-gen-at5}
\end{table}

We evaluate workaround behavior on a \emph{canonically filtered} subset of the Generalization set. For each problem, we first identify the target API from the benchmark-provided canonical solution, then retain only those problems for which the pre-edit model generation also uses that API. This ensures that any post-edit success without the edited API reflects a true workaround rather than a case where the model never relied on the target API to begin with. Table~\ref{tab:canonical-instant-gen-at5} decomposes Pass@5 into \emph{Successful Adoption} (A@5) and \emph{Workaround} (W@5), allowing us to distinguish true API migration from bypassing behavior.

Across all models and methods, a substantial portion of apparent success arises from workarounds rather than true adoption. For example, on CodeLlama, FT-L achieves Pass@5 of 0.40, but only 0.17 corresponds to successful adoption, while 0.23 results from workarounds. GRACE shows an even larger gap, with 0.41 Pass@5 but only 0.13 adoption and 0.28 workaround. Similar patterns hold across models. On CodeQwen, GRACE achieves 0.48 Pass@5, of which only 0.17 reflects adoption and 0.31 reflects workarounds. On DeepSeek, GRACE achieves 0.38 Pass@5, but only 0.19 corresponds to adoption, with 0.19 due to workarounds.

This gap is especially pronounced for locate-then-edit methods. MEMIT, PMET, and ROME often achieve moderate Pass@5 scores while exhibiting low adoption rates. For instance, on CodeQwen, PMET reaches 0.44 Pass@5, but only 0.16 corresponds to adoption, with the majority of success arising from workarounds (0.29). On DeepSeek, MEMIT and ROME show similarly low adoption rates despite non-zero Pass@5, indicating that these methods frequently fail to incorporate the edited API and instead rely on alternative implementations that satisfy the tests.

Even for the strongest methods, successful adoption remains substantially below overall task success. Across models, A@5 is often substantially smaller than Pass@5, indicating that many successful outputs do not reflect true API migration.

\subsection{Non-Workaroundable APIs}
\label{sec:non-workaroundable}
\begin{table}[t]
\centering
\scriptsize
\caption{Performance for single edits on the non-workaroundable subset. Only solutions using the target API can succeed.}
\vspace{-0.5em}
\setlength{\tabcolsep}{5pt}
\begin{tabular}{l|cc|cc|cc}
\toprule
& \multicolumn{2}{c|}{\textbf{CodeLlama}} & \multicolumn{2}{c|}{\textbf{CodeQwen}} & \multicolumn{2}{c}{\textbf{DeepSeek}} \\
\cmidrule(lr){2-3} \cmidrule(lr){4-5} \cmidrule(lr){6-7}
\textbf{Method} & Compiles@5 & Pass@5 & Compiles@5 & Pass@5 & Compiles@5 & Pass@5 \\
\midrule
pre-edit & 1.00 & 0.40 & 1.00 & 0.70 & 1.00 & 0.80 \\
\midrule
FT-L & 0.30 & 0.20 & 0.40 & 0.20 & 0.30 & 0.20 \\
GRACE & 0.20 & 0.20 & 0.30 & 0.10 & 0.40 & 0.20 \\
A-GRACE & 0.20 & 0.20 & 0.20 & 0.10 & 0.40 & 0.20 \\
MALMEN & 0.20 & 0.20 & 0.70 & 0.00 & 0.40 & 0.20 \\
MEMIT & 0.00 & 0.00 & 0.40 & 0.00 & 0.00 & 0.00 \\
PMET & 0.30 & 0.10 & 0.40 & 0.10 & 0.50 & 0.00 \\
ROME & 0.50 & 0.20 & 0.20 & 0.20 & 0.40 & 0.10 \\
\bottomrule
\end{tabular}
\vspace{-1em}
\label{tab:non-workaroundable-apis_at5}
\end{table}

To isolate true API adoption, we conduct a focused experiment on a new dataset of problems that strictly require the target API, eliminating workarounds. In this setting, passing behavior necessarily reflects correct use of the edited interface rather than workaround-based success. Construction details for this dataset are provided in Appendix~\ref{app:noworkaround}. Table~\ref{tab:non-workaroundable-apis_at5} reports model performance on this non-workaroundable set.

Performance drops sharply across all methods when workarounds are removed. On CodeLlama, Pass@5 reaches at most 0.20 for every method. The same upper bound holds for CodeQwen and DeepSeek, and several method-model pairs collapse to 0.00. This pattern is especially severe for locate-then-edit methods. MEMIT achieves 0.00 Pass@5 on all three models, while PMET and ROME frequently fall to 0.00 or 0.10. Even the stronger methods from earlier sections, including FT-L, GRACE, and A-GRACE, do not exceed 0.20 Pass@5 on any model. These results show that once workaround paths are eliminated, successful editing becomes rare.

The gap between Compiles@5 and Pass@5 further clarifies the failure mode. Many methods still produce compilable code, but this rarely translates into correct behavior. For example, MALMEN reaches Compiles@5 of 0.70 on CodeQwen while achieving 0.00 Pass@5, and PMET reaches Compiles@5 of 0.50 on DeepSeek with 0.00 Pass@5. 
More broadly, Compiles@5 remains non-zero for many methods across models, while Pass@5 often stays at or near zero. This indicates that the core challenge is not just producing syntactically valid code, but generating code that correctly realizes the edited API behavior.
\subsection{Successive Edits}
\label{sec:successive}

We now evaluate editing methods under successive edits, where multiple API modifications are applied cumulatively. This setting tests whether methods remain stable as edits accumulate. Table~\ref{tab:seq-results-gen-spec-agg} reports performance on both Generalization and Specificity.

Performance degrades sharply under successive edits. On the Generalization subset, Pass@5 falls from a baseline of 1.00 to near-zero for most methods. FT-L retains limited performance, reaching 0.26--0.42 across models, while GRACE remains the strongest method at 0.52--0.58. In contrast, A-GRACE degrades sharply, and MALMEN, MEMIT, PMET, and ROME collapse almost entirely, with Pass@5 dropping to 0.00 in most settings. The aggregate $\Delta P$ values confirm this trend, reaching as low as $-1.00$ for several methods.

A similar pattern appears on the Specificity set. While GRACE maintains strong performance, achieving 0.81--0.95 Pass@5, most other methods suffer severe degradation. FT-L drops to 0.33--0.40, A-GRACE collapses to 0.00 on two models, and MALMEN, MEMIT, PMET, and ROME again exhibit near-complete failure. These results indicate that successive edits not only impair adaptation to updated APIs but also introduce substantial collateral damage to unrelated tasks.

Successive edits often trigger model-specific collapse modes. PMET exhibits severe generation instability. On DeepSeek, editing collapses entirely, producing no usable output. On CodeQwen, collapse yields repetitive \texttt{defdefdef...} token sequences that may compile as valid identifiers but fail at runtime, or degenerate into syntax errors when gaps appear between tokens. 
On CodeLlama, PMET instead produces malformed repetitive code fragments that are rejected as syntax errors. MALMEN on CodeQwen and DeepSeek shows a different failure mode, collapsing to garbage token sequences that fail to compile. On CodeLlama, some of these sequences still compile because Python interprets them as single valid identifiers, yet they fail all tests. 
These examples suggest that successive edits not only reduce task accuracy but also destabilize the generation process. 

Shapley attribution shows that compilation errors increasingly drive these failures. On the Generalization subset, CF rises sharply for most methods, reaching 73.15\% for MALMEN and 100\% for MEMIT. Similar trends appear on Specificity, where MEMIT and ROME exceed 80\% CF. In contrast, GRACE maintains low CF on Specificity (4.92\%), indicating better preservation of syntactic stability under accumulative edits.

\begin{table*}[t]
\centering
\scriptsize
\caption{Performance under successive edits on Generalization and Specificity. Aggregate $\Delta P$ and CF are averaged across models.}
\vspace{-1em}
\setlength{\tabcolsep}{2.5pt}
\begin{tabular}{l|cc|cc|cc|cc|cc|cc|cc|cc}
\toprule
& \multicolumn{8}{c|}{\textbf{Generalization}} & \multicolumn{8}{c}{\textbf{Specificity}} \\
\cmidrule(lr){2-9} \cmidrule(lr){10-17}
& \multicolumn{2}{c|}{\textbf{CodeLlama}} & \multicolumn{2}{c|}{\textbf{CodeQwen}} & \multicolumn{2}{c|}{\textbf{DeepSeek}} & \multicolumn{2}{c|}{\textbf{Average}} & \multicolumn{2}{c|}{\textbf{CodeLlama}} & \multicolumn{2}{c|}{\textbf{CodeQwen}} & \multicolumn{2}{c|}{\textbf{DeepSeek}} & \multicolumn{2}{c}{\textbf{Average}} \\
\cmidrule(lr){2-3} \cmidrule(lr){4-5} \cmidrule(lr){6-7} \cmidrule(lr){8-9}
\cmidrule(lr){10-11} \cmidrule(lr){12-13} \cmidrule(lr){14-15} \cmidrule(lr){16-17}
\textbf{Method} & C@5 & P@5 & C@5 & P@5 & C@5 & P@5 & $\Delta$P & CF & C@5 & P@5 & C@5 & P@5 & C@5 & P@5 & $\Delta$P & CF \\
\midrule
pre-edit & 1.00 & 1.00 & 1.00 & 1.00 & 1.00 & 1.00 & -- & -- & 1.00 & 1.00 & 1.00 & 1.00 & 1.00 & 1.00 & -- & -- \\
\midrule
FT-L & 0.76 & 0.26 & 0.88 & 0.42 & 0.54 & 0.29 & -0.68 & 28.97 & 0.93 & 0.40 & 0.99 & 0.35 & 0.57 & 0.33 & -0.64 & 19.48 \\
GRACE & 0.63 & 0.52 & 0.67 & 0.58 & 0.70 & 0.55 & -0.45 & 67.10 & 1.00 & 0.92 & 1.00 & 0.95 & 0.97 & 0.81 & -0.11 & 4.92 \\
A-GRACE & 0.04 & 0.02 & 0.72 & 0.61 & 0.02 & 0.02 & -0.78 & 80.42 & 0.00 & 0.00 & 1.00 & 0.95 & 0.00 & 0.00 & -0.68 & 66.67 \\
MALMEN & 0.61 & 0.00 & 0.00 & 0.00 & 0.00 & 0.00 & -1.00 & 73.15 & 0.53 & 0.00 & 0.00 & 0.00 & 0.00 & 0.00 & -1.00 & 74.42 \\
MEMIT & 0.00 & 0.00 & 0.00 & 0.00 & 0.00 & 0.00 & -1.00 & 100.00 & 0.00 & 0.00 & 0.00 & 0.00 & 0.00 & 0.00 & -1.00 & 100.00 \\
PMET & 0.00 & 0.00 & 1.00 & 0.00 & 0.00 & 0.00 & -1.00 & 66.67 & 0.00 & 0.00 & 1.00 & 0.00 & 0.00 & 0.00 & -1.00 & 66.67 \\
ROME & 0.18 & 0.00 & 0.00 & 0.00 & 0.00 & 0.00 & -1.00 & 80.28 & 0.15 & 0.00 & 0.00 & 0.00 & 0.00 & 0.00 & -1.00 & 80.83 \\
\bottomrule
\end{tabular}
\vspace{-2em}
\label{tab:seq-results-gen-spec-agg}
\end{table*}
\section{Discussion}

Code editing differs from traditional natural-language fact editing because code generation admits multiple solution possibilities that can satisfy test cases without using the edited API. As a result, aggregate metrics alone can be misleading. Our results show that many passing outputs reflect workaround behavior rather than genuine API adoption, with successful adoption often substantially smaller than Pass@k on the canonically filtered Generalization set. This suggests that evaluation of API editing should go beyond task success and explicitly account for whether the edited interface is actually used.

These outcomes are best interpreted through the stage-based taxonomy in Table~\ref{tab:taxonomy}. The taxonomy separates \emph{Successful Adoption} from \emph{Workaround Success}, and further partitions failures into \emph{Compilation}, \emph{API}, \emph{Execution}, and \emph{Behavioral} stages. This organization highlights two key properties of code editing. First, passing tests may reflect either correct API usage or workaround behavior. Second, non-passing outputs are not uniform, but fail at different stages, ranging from syntax and malformed generation to incorrect interface use, runtime exceptions, and incorrect behavior after successful execution. API failures are particularly important in our setting because they isolate cases where the model produces code that may still compile, yet fails to realize the edited interface. These include retaining the old API, invoking the updated API incorrectly, fabricating nonexistent interfaces, and failing to adapt to changed return types. Such distinctions are obscured by aggregate metrics alone, but are central to understanding how editing succeeds or fails in code generation. 

These results also show that failure becomes more severe and qualitatively different across evaluation settings. Under single edits, degradation reflects both compilation and post-compilation failures, showing that editing can disrupt both syntactic validity and correct API adaptation (Section~\ref{sec:single-edit}). Moreover, much of the apparent success in this setting comes from workaround-based solutions rather than genuine use of the edited API (Section~\ref{sec:workarounds}). When workaround possibilities are removed, performance drops sharply even when compilability remains non-zero, indicating that runnable code is often insufficient to realize the intended API behavior (Section~\ref{sec:non-workaroundable}). Under successive edits, degradation becomes more severe and increasingly concentrates at compilation, with many methods collapsing through malformed generation, repetitive fragments, or invalid decoding~(Section~\ref{sec:successive}). These patterns show that accumulated edits not only reduce overall task performance but also increasingly destabilize code generation at the syntactic level.

These findings have implications for method design. The compilation vs post-compilation contrast revealed by our Shapley analysis suggests that editing methods should not optimize only for aggregate Pass@k. Methods that fail at compilation break the model’s ability to produce syntactically valid and interface-compatible code, whereas methods that fail after compilation can still generate runnable programs but do not correctly realize the edited behavior. Future work should therefore target these failure modes separately rather than treating them as a single notion of error.

Overall, current editing methods are not yet suitable for realistic API migration. Even when single edits achieve moderate task success, much of that success is workaround-based rather than true adoption. When workaround possibilities are removed, performance drops sharply (Section~\ref{sec:non-workaroundable}), and under successive edits, most methods become unstable or collapse entirely (Section~\ref{sec:successive}). Progress will require not only more accurate editing methods but also evaluation metrics, benchmark designs, and model architectures that support reliable, compositional updates in evolving software ecosystems.

\begin{table}[t]
\centering
\footnotesize
\caption{Taxonomy of post-edit outcomes based on execution stages, separating adoption, workaround success, and failures across compilation, API usage, execution, and behavior.}
\vspace{-0.5em}
\label{tab:taxonomy}

\begin{tabular}{p{3cm} p{2.3cm} p{6.9cm}}
\toprule
\textbf{Outcome} & \textbf{Subtype} & \textbf{Description} \\
\midrule

\textbf{Successful Adoption} 
& -- 
& Code passes tests and uses the new API \\

\textbf{Workaround Success} 
& -- 
& Code passes tests without using the new API, relying on alternatives \\

\midrule

\multirow{3}{*}{\textbf{Compilation Failure}}
& Syntax Error 
& Code fails to compile because of invalid syntax, such as missing keywords, indentation errors, or malformed expressions \\

& Degenerate Generation 
& Code fails to compile because generation collapses into repetitive fragments or incomplete code \\

& Invalid Definition 
& Code fails to compile because of a malformed function signature or inconsistent definition structure \\

\midrule

\multirow{4}{*}{\textbf{API Failure}}
& Old API Usage 
& Code uses the old API instead of the new interface \\

& Incorrect Edited Invocation 
& Code uses the edited API incorrectly, such as with missing required arguments, wrong argument order, or invalid parameter names \\

& Fabricated API 
& Code generates a plausible yet non-existent function name or interface \\

& Return/Type Mismatch 
& Code fails to adapt to a modified return schema or expected data type \\

\midrule

\textbf{Execution Failure}
& Exception 
& Code compiles but raises exception(s) at runtime, whether it uses the updated API or workarounds\\

\midrule

\textbf{Behavioral Failure}
& Test Failure 
& Code compiles and executes without exception but fails one or more unit tests,  whether it uses the updated API or workarounds \\

\bottomrule
\end{tabular}
\vspace{-1em}
\end{table}


\section{Conclusion}

We studied API-update editing in code LLMs and found that current methods do not reliably induce correct API migration. Much of their apparent success arises from workaround-based solutions rather than true adoption of updated interfaces, and when workaround paths are removed or edits accumulate, performance drops sharply and often collapses. Our analysis further shows that failures are not uniform but arise at different stages, including compilation and post-compilation behavior, highlighting the need for evaluation beyond Pass@$k$ with explicit measurement of API adoption and failure modes. Overall, current editing approaches are not yet robust to evolving APIs, and progress will require methods and evaluation frameworks that support reliable, compositional updates in code generation.
  
\bibliography{colm2026_conference}
\bibliographystyle{colm2026_conference}
\clearpage
\appendix
\onecolumn
\section{Dataset Construction Details}
\label{app:dataset}

\subsection{Preprocessing Pipeline}
From APPS' 232,421 candidate solutions across 10,000 competition-level problems, we sample up to 5 solutions per problem and retain only function-based implementations (standalone functions and class methods), yielding 12,892 testable solutions. We execute each solution against its unit tests, filtering to 10,056 solutions that pass all tests (78\% pass rate). From these verified solutions, we extract API usage patterns, identifying 213 unique APIs spanning Python's standard library (e.g., \texttt{collections.Counter}, \texttt{itertools.combinations}) and built-in functions (e.g., \texttt{sorted}, \texttt{enumerate}), yielding 20,015 problem-API pairs. We integrate these with MBPP (974 problems, 1,156 pairs) and HumanEval (164 problems, 259 pairs), remove external libraries and the \texttt{print} function, resulting in 21,430 problem-API pairs from 11,138 problems.

\subsection{Sampling Strategy}
\label{app:sampling}
From 21,430 problem-API pairs, we strategically sample 2,040 instances in a 1:2:2 ratio across Reliability, Generalization, and Specificity. For Specificity (target: 1,000), we take all problems with rare APIs and fill the remaining budget by evenly sampling from the top most frequent APIs. For Reliability and Generalization, we use stratified 1:2 assignment: for each API, we shuffle all problems using that API and assign the first third to Reliability and remaining two-thirds to Generalization, ensuring both splits share identical API coverage while maintaining disjoint problem sets. We then apply proportional downsampling, allocating samples to each API proportional to its usage frequency (minimum 2 per API), to reach targets of 500 (Reliability) and 1,000 (Generalization).
We initially targeted a 500/1000/1000 split, but after applying the final filtering, coverage, and downsampling constraints, the realized split sizes were 469, 647, and 924 instances for Reliability, Generalization, and Specificity, respectively.Construction
\section{Non-Workaroundable Dataset Construction}
\label{app:noworkaround}
To study model behavior post-editing when workarounds are unavailable, we construct a small dataset of 5 editing instances comprising 5 reliability problems and 10 generalization problems, where each editing instance is paired with 2 generalization problems requiring use of the same API on distinct tasks, yielding 15 problems in total. Selecting genuinely non-workaroundable APIs in Python is non-trivial, as the ecosystem offers numerous alternative packages and equivalent low-level calls for most operations. We initially targeted OS-level and hardware-level APIs such as \texttt{os.getpid()} and \texttt{torch.cuda.device\_count()}, but found that workarounds remained feasible even for these. To eliminate workaround paths, we imposed two additional constraints during problem construction: (1) the problem prompt explicitly requires use of the target library, and (2) we manually verified that no alternative within that library can accomplish the same task. This process yielded 5 editing instances spanning all five modification types: R1 (\texttt{os.getpid}), S1 (\texttt{torch.cuda.device\_count}), S2 (\texttt{sqlite3.connect}), S3 (\texttt{shutil.move}), and S4 (\texttt{sys.getrecursionlimit}).
\section{Editing Methods Additional Details}
\label{app:methods}
\textbf{Fine-Tuning (FT-L).}

Fine-tuning~\citep{ft} updates a targeted subset of model parameters while freezing the remainder, a design choice motivated by the small scale of typical editing datasets. It minimizes prediction error on the target editing examples while applying norm constraints on parameter updates to limit interference with unrelated knowledge.
    
\textbf{GRACE.}~\citet{hartvigsen2023aging} employs an external memory mechanism that bypasses direct parameter updates by maintaining a codebook of edited knowledge alongside the frozen base model. This memorization-based approach stores edits as key-value pairs: keys are generated from the model's hidden state activations at a designated layer when processing inputs requiring modification, while values are learned vectors encoding the corrected information through backpropagation over the editing objective. During inference, when an input produces a hidden state matching a stored key, the associated value substitutes the original activation for subsequent layers, redirecting the model's output without altering its weights. Though this strategy guarantees preservation of the base model's original knowledge, it incurs additional computational cost from memory retrieval operations at inference time.

\textbf{A-GRACE.}~\citet{10.1109/ICSE55347.2025.00049} extends GRACE with a two-layer MLP encoder that averages token representations as keys, improving generalization. The encoder is trained contrastively to minimize distance between reliability and generalization prompts while maximizing distance from specificity prompts.

\textbf{MALMEN.}~\citet{tan2024massiveeditinglargelanguage} achieves scalable multi-fact editing by addressing two key challenges in hyper-network methods. First, simultaneous edits often require conflicting parameter changes. Existing hyper-networks generate individual parameter shifts and sum them together (analogous to gradient accumulation), causing cancellation effects where contradictory updates neutralize each other. MALMEN instead formulates parameter shift aggregation as a least squares problem, finding a single optimal shift that best satisfies all edits. Second, standard hyper-networks create memory bottlenecks by processing all edits through the full LM in a single forward pass. MALMEN decouples hyper-network computation from the LM, enabling independent batch sizes that dramatically reduce memory consumption. 

\textbf{Rank-One Model Editing (ROME).}~\citet{meng2022locating} employs localized parameter modification, targeting specific weights rather than applying global updates. The method uses causal tracing~\citep{NEURIPS2020_92650b2e} to identify critical layers where factual knowledge resides, operating under the principle that Feed-Forward Networks (FFNs) function as associative key-value memories for storing facts~\citep{geva-etal-2021-transformer}. ROME formulates editing as a constrained least-squares optimization problem, directly updating FFN weight matrices to produce correct outputs for modified facts without requiring iterative gradient descent. This closed-form solution enables computationally efficient single-fact corrections. However, ROME's architecture restricts it to processing one edit at a time, preventing batch or sequential editing. This constrains its utility for scenarios requiring continuous knowledge updates, though it excels at precise, isolated corrections.    

\textbf{MEMIT.}~\citet{meng2023memit} extends ROME's rank-one approach to accommodate batch edits by targeting multiple layers simultaneously. While ROME modifies a single critical layer, MEMIT uses causal tracing to identify a range of mediating layers that collectively encode factual knowledge about a subject, then distributes weight updates across these layers' MLP components. This multi-layer intervention is designed such that by the final mediating layer, the model's hidden states reflect all inserted memories. The distributed update strategy enables MEMIT to scale to substantially larger edit batches than ROME's single-edit constraint allows. However, MEMIT exhibits instability during sequential editing~\citep{gupta-etal-2024-model}, limiting its effectiveness for continuous knowledge maintenance scenarios.

\textbf{PMET.}~\citet{li2024pmet} addresses a key limitation in ROME and MEMIT, which use Transformer Layer (TL) hidden states as knowledge representations for updating FFN weights. Since TL states aggregate information from Multi-Head Self-Attention (MHSA), FFN, and residual connections, they introduce irrelevant information that degrades editing precision. Through analysis of component-level representations, PMET observes that MHSA encodes general knowledge extraction patterns rather than specific facts, suggesting MHSA weights need not be modified when inserting new factual knowledge. Leveraging this insight, PMET optimizes MHSA and FFN hidden states separately but uses only the refined FFN representations as targets for weight updates, eliminating contamination from MHSA information and enabling more precise parameter modifications.
\section{Method Configurations.}
\label{app:methods_config}
We configure editing methods following established protocols~\citep{wang-etal-2024-easyedit, 10.1109/ICSE55347.2025.00049}. \textit{FT-L} fine-tunes the down-projection matrix of layer 21 for 25 iterations with learning rate $5\times10^{-4}$ and $L_\infty$-norm constraint $\delta = 1\times10^{-4}$. \textit{GRACE} inserts an adapter at layer 27 with key radius 1, optimizing value vectors for 30 iterations at learning rate 1, evaluated under budgets of 30, 90, and 270 edits. \textit{ROME} targets layer 5, while \textit{MEMIT} and \textit{PMET} modify layers \{4, 5, 6, 7, 8\}. Layer selections follow prior work~\citep{hase2023does, wang-etal-2024-easyedit}, as causal tracing shows no correlation with editing success~\citep{hase2023does}. \textit{MALMEN} trains a two-layer MLP hypernetwork to predict parameter shifts for the down-projection matrices of the last 5 layers, with a middle layer dimension of 1920, learning rate $1\times10^{-5}$, batch size 4, and a maximum of 1,000 training steps, with hypernetwork parameters clamped to $L_2$ norm $\leq 1$. \textit{A-GRACE} augments GRACE with a two-layer MLP encoder (output dimension 256) trained with a margin of 0.5. The encoder is trained for up to 100 epochs with batch size 64, learning rate $1\times10^{-4}$, weight decay $1\times10^{-4}$, and early stopping with patience 10.
\section{Shapley Attribution}
\label{app:shapley}

\subsection{Shapley Attribution Derivation}
\label{app:shapley_der}

To understand \emph{why} an editing method changes end-to-end success, we decompose Pass@$k$ into two interpretable factors: (i) the ability to produce at least one \emph{compilable} candidate among the top-$k$ samples, and (ii) correctness \emph{given} that a compilable candidate exists. Formally, let $C = \texttt{Compiles@k}$, $P = \texttt{Pass@k}$, and $Q = P(\text{Pass} \mid \text{Compiled})$. Since any passing program must compile, $\text{Pass} \subseteq \text{Compiled}$, which gives $Q = P/C$ (with $Q=0$ when $C=0$), yielding the identity:

\begin{equation}
P = C \cdot Q.
\end{equation}

We compare an edited method (subscript $1$) to the non-edit baseline (subscript $0$), with total change $\Delta P = P_1 - P_0$. Because $P$ is the product of $C$ and $Q$, changes in compilability and post-compilation correctness interact, making it ambiguous to attribute $\Delta P$ to one factor alone. To resolve this, we apply an exact \textit{two-factor Shapley attribution}, which averages marginal contributions over the two possible orders of applying the changes, yielding closed-form contributions:

\begin{equation}
\phi_C = (C_1 - C_0)\cdot \frac{Q_0 + Q_1}{2}, \quad
\phi_Q = (Q_1 - Q_0)\cdot \frac{C_0 + C_1}{2}
\end{equation}

such that $\Delta P = \phi_C + \phi_Q$ (up to rounding). Intuitively, $\phi_C$ measures how much the change in compilation probability affects success evaluated at the average post-compilation correctness, while $\phi_Q$ measures how much the change in post-compilation correctness affects success evaluated at the average compilation probability. 
To attribute the performance change, $\Delta P$, we compute the compilation failure~(CF) contribution:

\begin{equation}
\text{CF} = \frac{|\phi_C|}{|\phi_C| + |\phi_Q|} \times 100
\end{equation}

which lies in $[0,100]$. Values close to $100$ indicate the performance drop is primarily driven by \emph{loss of compilable outputs}, whereas values close to $0$ indicate the drop is primarily due to \emph{post-compilation failures} among otherwise compilable candidates.

\clearpage
\subsection{Single Edit Shapley Two-Factor Contribution}

Tables~\ref{tab:results-codellama-shapley-instant}, \ref{tab:results-codeqwen-shapley-instant}, and~\ref{tab:results-deepseek-shapley-instant} present the per-model Shapley decompositions underlying the aggregate results in Table~\ref{tab:instant-results-gen-spec-at5}. For each model and method, we report the total performance drop ($\Delta$P), the compilability contribution ($\phi_C$), the post-compilation correctness contribution ($\phi_Q$), and CF, which measures the fraction of the drop attributable to loss of compilable outputs.

\begin{table*}[h!]
\centering
\footnotesize
\begin{tabular}{l|r|r|r|r|r|r|r|r}
\toprule

\multirow{2}{*}{\textbf{Method}} & \multicolumn{4}{c|}{Generalization} & \multicolumn{4}{c}{Specificity} \\
\cmidrule(lr){2-5} \cmidrule(lr){6-9}
& $\Delta$P & $\phi$C & $\phi$Q & CF & $\Delta$P & $\phi$C & $\phi$Q & CF \\
\midrule
FT-L & -0.37 & -0.22 & -0.14 & 61.28 & -0.12 & 0.00 & -0.12 & 0.00 \\
GRACE & -0.36 & -0.24 & -0.12 & 66.22 & -0.08 & 0.00 & -0.08 & 0.00 \\
A-GRACE & -0.38 & -0.23 & -0.15 & 60.49 & -0.10 & -0.00 & -0.10 & 4.57 \\
MALMEN & -0.56 & -0.23 & -0.33 & 41.26 & -0.32 & -0.04 & -0.27 & 13.22 \\
MEMIT & -0.74 & -0.38 & -0.36 & 51.65 & -0.63 & -0.26 & -0.37 & 41.96 \\
PMET & -0.68 & -0.29 & -0.40 & 41.79 & -0.53 & -0.17 & -0.37 & 31.05 \\
ROME & -0.65 & -0.26 & -0.39 & 39.76 & -0.55 & -0.10 & -0.45 & 18.16 \\
\bottomrule
\end{tabular}
\caption{Single Edit Results for \textbf{CodeLlama} across Generalization and Specificity. $\Delta$P = performance drop, $\phi$C and $\phi$Q are model-level contributions, CF is the compilation contribution.}
\label{tab:results-codellama-shapley-instant}
\end{table*}

\textbf{CodeLlama.} Compilation-driven failure is the dominant factor on Generalization for all methods (CF = 39--66\%). On Specificity, CF is near zero for FT-L, GRACE, and A-GRACE, and remains comparatively small for the remaining methods, indicating that across the board, Specificity drops are driven primarily by post-compilation correctness failures rather than loss of compilable outputs.

\begin{table*}[h!]
\centering
\footnotesize
\begin{tabular}{l|r|r|r|r|r|r|r|r}
\toprule

\multirow{2}{*}{\textbf{Method}} & \multicolumn{4}{c|}{Generalization} & \multicolumn{4}{c}{Specificity} \\
\cmidrule(lr){2-5} \cmidrule(lr){6-9}
& $\Delta$P & $\phi$C & $\phi$Q & CF & $\Delta$P & $\phi$C & $\phi$Q & CF \\
\midrule
FT-L & -0.32 & -0.22 & -0.10 & 70.12 & -0.08 & 0.00 & -0.08 & 0.00 \\
GRACE & -0.31 & -0.22 & -0.09 & 69.85 & -0.07 & 0.00 & -0.07 & 0.00 \\
A-GRACE & -0.34 & -0.22 & -0.12 & 65.38 & -0.06 & -0.00 & -0.06 & 3.91 \\
MALMEN & -0.49 & -0.23 & -0.26 & 46.66 & -0.33 & -0.08 & -0.25 & 23.70 \\
MEMIT & -0.42 & -0.24 & -0.18 & 57.44 & -0.19 & -0.05 & -0.14 & 26.14 \\
PMET & -0.35 & -0.24 & -0.11 & 69.69 & -0.09 & -0.02 & -0.07 & 21.50 \\
ROME & -0.45 & -0.23 & -0.22 & 51.51 & -0.25 & -0.04 & -0.21 & 16.28 \\
\bottomrule
\end{tabular}
\caption{Single Edit Results for \textbf{CodeQwen} across Generalization and Specificity. $\Delta$P = performance drop, $\phi$C and $\phi$Q are model-level contributions, CF is the compilation contribution.}
\label{tab:results-codeqwen-shapley-instant}
\end{table*}
\textbf{CodeQwen.} CodeQwen consistently exhibits the lowest $\Delta$P across all methods compared to CodeLlama and DeepSeek, on both Generalization and Specificity, with $\Delta$P ranging from $-0.49$ to $-0.31$ on Generalization and $-0.33$ to $-0.06$ on Specificity. Also worth noting is that for CodeQwen, the drop in Pass@k on Generalization is driven predominantly by compilation failures (CF = 47--70\%), whereas on Specificity the drop is driven predominantly by post-compilation correctness failures (CF = 0--26\%), consistent with the overall pattern observed across models.

\clearpage

\begin{table*}[h!]
\centering
\footnotesize
\begin{tabular}{l|r|r|r|r|r|r|r|r}
\toprule

\multirow{2}{*}{\textbf{Method}} & \multicolumn{4}{c|}{Generalization} & \multicolumn{4}{c}{Specificity} \\
\cmidrule(lr){2-5} \cmidrule(lr){6-9}
& $\Delta$P & $\phi$C & $\phi$Q & CF & $\Delta$P & $\phi$C & $\phi$Q & CF \\
\midrule
FT-L & -0.34 & -0.16 & -0.17 & 48.51 & -0.20 & -0.03 & -0.17 & 17.17 \\
GRACE & -0.34 & -0.15 & -0.19 & 43.72 & -0.17 & -0.05 & -0.12 & 27.77 \\
A-GRACE & -0.36 & -0.17 & -0.18 & 48.03 & -0.18 & -0.05 & -0.13 & 28.96 \\
MALMEN & -0.43 & -0.18 & -0.25 & 41.31 & -0.39 & -0.09 & -0.30 & 23.09 \\
MEMIT & -0.77 & -0.42 & -0.35 & 54.70 & -0.72 & -0.38 & -0.34 & 52.52 \\
PMET & -0.75 & -0.39 & -0.37 & 51.38 & -0.63 & -0.32 & -0.31 & 51.16 \\
ROME & -0.74 & -0.47 & -0.28 & 62.53 & -0.64 & -0.35 & -0.29 & 54.46 \\
\bottomrule
\end{tabular}
\caption{Single Edit Results for \textbf{DeepSeek} across Generalization and Specificity. $\Delta$P = performance drop, $\phi$C and $\phi$Q are model-level contributions, CF is the compilation contribution.}
\label{tab:results-deepseek-shapley-instant}
\end{table*}
\textbf{DeepSeek.} The most distinctive pattern for DeepSeek is on Specificity, where CF is consistently higher across all methods compared to CodeLlama and CodeQwen, including for FT-L, GRACE, and A-GRACE (17--29\% vs. near zero for the other models).

It can be seen from Tables~\ref{tab:results-codellama-shapley-instant}, \ref{tab:results-codeqwen-shapley-instant}, and~\ref{tab:results-deepseek-shapley-instant} that CF is consistently higher on Generalization than Specificity, confirming that LLMs produce more structurally broken code for problems that share the same API as the edited problem than for problems involving unmodified APIs.

\subsection{Successive Edits Shapley Two-Factor Contribution}

Tables~\ref{tab:results-codellama-shapley-seq}, \ref{tab:results-codeqwen-shapley-seq}, and~\ref{tab:results-deepseek-shapley-seq} present the per-model Shapley decompositions underlying the aggregate results in Table~\ref{tab:seq-results-gen-spec-agg}. The most striking difference from the single edit setting is that the majority of methods reach $\Delta$P = $-$1.00, indicating complete performance collapse. The decomposition here is therefore most informative for understanding how methods collapse: whether through total loss of compilable outputs (CF $\approx$100\%) or through runtime failures among otherwise compilable outputs (CF $\approx$0\%).

\begin{table*}[h!]
\centering
\footnotesize
\begin{tabular}{l|r|r|r|r|r|r|r|r}
\toprule
\multirow{2}{*}{\textbf{Method}} & \multicolumn{4}{c|}{Generalization} & \multicolumn{4}{c}{Specificity} \\
\cmidrule(lr){2-5} \cmidrule(lr){6-9}
& $\Delta$P & $\phi$C & $\phi$Q & CF & $\Delta$P & $\phi$C & $\phi$Q & CF \\
\midrule
FT-L & -0.74 & -0.16 & -0.58 & 21.95 & -0.60 & -0.05 & -0.55 & 8.64 \\
GRACE & -0.48 & -0.33 & -0.14 & 70.28 & -0.08 & 0.00 & -0.08 & 0.00 \\
A-GRACE & -0.98 & -0.72 & -0.26 & 73.59 & -1.00 & -1.00 & 0.00 & 100.00 \\
MALMEN & -1.00 & -0.19 & -0.81 & 19.45 & -1.00 & -0.23 & -0.77 & 23.26 \\
MEMIT & -1.00 & -1.00 & 0.00 & 100.00 & -1.00 & -1.00 & 0.00 & 100.00 \\
PMET & -1.00 & -1.00 & 0.00 & 100.00 & -1.00 & -1.00 & 0.00 & 100.00 \\
ROME & -1.00 & -0.41 & -0.59 & 40.85 & -1.00 & -0.42 & -0.58 & 42.48 \\
\bottomrule
\end{tabular}
\caption{Successive Edits Results for \textbf{CodeLlama} across Generalization and Specificity. $\Delta$P = performance drop, $\phi$C and $\phi$Q are model-level contributions, CF is the compilation contribution.}
\label{tab:results-codellama-shapley-seq}
\end{table*}

\textbf{CodeLlama.} GRACE and FT-L are the only methods to avoid full collapse. GRACE achieves $\Delta$P = $-$0.48 on Generalization and $-$0.08 on Specificity, with Generalization failures driven predominantly by compilation loss (CF = 70.28\%). FT-L shows larger drops ($\Delta$P = $-$0.74 on Generalization, $-$0.60 on Specificity) with failures driven almost entirely by runtime failures on both splits (CF = 21.95\% and 8.64\%). A-GRACE near-fully collapses on Generalization ($\Delta$P = $-$0.98, CF = 73.59\%) and fully collapses on Specificity (CF = 100\%), producing no compilable outputs. MEMIT and PMET fully collapse on both splits with CF = 100\%, similarly producing no compilable outputs. MALMEN and ROME also fully collapse on both splits but with lower CF (19--23\% and 41--42\% respectively), indicating that compilable outputs are still generated but fail at runtime.

\begin{table*}[h!]
\centering
\footnotesize
\begin{tabular}{l|r|r|r|r|r|r|r|r}
\toprule
\multirow{2}{*}{\textbf{Method}} & \multicolumn{4}{c|}{Generalization} & \multicolumn{4}{c}{Specificity} \\
\cmidrule(lr){2-5} \cmidrule(lr){6-9}
& $\Delta$P & $\phi$C & $\phi$Q & CF & $\Delta$P & $\phi$C & $\phi$Q & CF \\
\midrule
FT-L & -0.58 & -0.09 & -0.49 & 14.88 & -0.65 & -0.00 & -0.65 & 0.52 \\
GRACE & -0.42 & -0.30 & -0.12 & 71.42 & -0.05 & 0.00 & -0.05 & 0.00 \\
A-GRACE & -0.39 & -0.26 & -0.12 & 67.67 & -0.05 & 0.00 & -0.05 & 0.00 \\
MALMEN & -1.00 & -1.00 & 0.00 & 100.00 & -1.00 & -1.00 & 0.00 & 100.00 \\
MEMIT & -1.00 & -1.00 & 0.00 & 100.00 & -1.00 & -1.00 & 0.00 & 100.00 \\
PMET & -1.00 & 0.00 & -1.00 & 0.00 & -1.00 & 0.00 & -1.00 & 0.00 \\
ROME & -1.00 & -1.00 & 0.00 & 100.00 & -1.00 & -1.00 & 0.00 & 100.00 \\
\bottomrule
\end{tabular}
\caption{Successive Edits Results for \textbf{CodeQwen} across Generalization and Specificity. $\Delta$P = performance drop, $\phi$C and $\phi$Q are model-level contributions, CF is the compilation contribution.}
\label{tab:results-codeqwen-shapley-seq}
\end{table*}

\textbf{CodeQwen.} GRACE and A-GRACE both avoid full collapse, with small Specificity drops ($\Delta$P = $-$0.05 for both) and moderate Generalization drops ($\Delta$P = $-$0.42 and $-$0.39 respectively). For both methods, Generalization failures are driven predominantly by compilation loss (CF = 71.42\% and 67.67\%), while Specificity failures are driven entirely by runtime failures (CF $\approx$0\%). FT-L partially survives but with larger drops ($\Delta$P = $-$0.58 on Generalization, $-$0.65 on Specificity), with failures driven almost entirely by runtime failures on both splits (CF = 14.88\% and 0.52\%). MALMEN, MEMIT, and ROME fully collapse on both splits with CF = 100\%, producing no compilable outputs. PMET is notably distinct: despite full collapse ($\Delta$P = $-$1.00 on both splits), CF = 0\% on both splits, meaning compilable outputs are still produced but all fail at runtime.

\begin{table*}[h!]
\centering
\footnotesize
\begin{tabular}{l|r|r|r|r|r|r|r|r}
\toprule
\multirow{2}{*}{\textbf{Method}} & \multicolumn{4}{c|}{Generalization} & \multicolumn{4}{c}{Specificity} \\
\cmidrule(lr){2-5} \cmidrule(lr){6-9}
& $\Delta$P & $\phi$C & $\phi$Q & CF & $\Delta$P & $\phi$C & $\phi$Q & CF \\
\midrule
FT-L & -0.71 & -0.36 & -0.35 & 50.09 & -0.68 & -0.33 & -0.34 & 49.28 \\
GRACE & -0.45 & -0.27 & -0.18 & 59.60 & -0.19 & -0.03 & -0.17 & 14.75 \\
A-GRACE & -0.98 & -0.98 & 0.00 & 100.00 & -1.00 & -1.00 & 0.00 & 100.00 \\
MALMEN & -1.00 & -1.00 & 0.00 & 100.00 & -1.00 & -1.00 & 0.00 & 100.00 \\
MEMIT & -1.00 & -1.00 & 0.00 & 100.00 & -1.00 & -1.00 & 0.00 & 100.00 \\
PMET & -1.00 & -1.00 & 0.00 & 100.00 & -1.00 & -1.00 & 0.00 & 100.00 \\
ROME & -1.00 & -1.00 & 0.00 & 100.00 & -1.00 & -1.00 & 0.00 & 100.00 \\
\bottomrule
\end{tabular}
\caption{Successive Edits Results for \textbf{DeepSeek} across Generalization and Specificity. $\Delta$P = performance drop, $\phi$C and $\phi$Q are model-level contributions, CF is the compilation contribution.}
\label{tab:results-deepseek-shapley-seq}
\end{table*}

\textbf{DeepSeek.} GRACE and FT-L are the only methods to avoid full collapse. GRACE achieves $\Delta$P = $-$0.45 on Generalization and $-$0.19 on Specificity, with Generalization failures driven predominantly by compilation loss (CF = 59.60\%) and Specificity failures driven predominantly by runtime failures (CF = 14.75\%). FT-L shows larger drops ($\Delta$P = $-$0.71 on Generalization, $-$0.68 on Specificity), with failures split roughly evenly between compilation loss and runtime failures on both splits (CF = 50.09\% and 49.28\%). A-GRACE near-fully collapses on Generalization ($\Delta$P = $-$0.98, CF = 100\%) and fully collapses on Specificity (CF = 100\%), producing no compilable outputs on either split. MALMEN, MEMIT, PMET, and ROME all fully collapse on both splits with CF = 100\%, producing no compilable outputs at all.

Across all three models, successive editing leads to near-universal performance collapse, with only GRACE and FT-L consistently avoiding full degradation. MEMIT is the most consistent in its failure mode, reaching full collapse with CF = 100\% across all three models, indicating total loss of compilable output regardless of the base model. For most other methods, however, the failure mode varies across models: MALMEN collapses with low CF on CodeLlama but with CF = 100\% on CodeQwen and DeepSeek, and similar inconsistencies are observed for ROME, PMET, and A-GRACE. This suggests that under successive edits, the dominant failure mode is not solely determined by the editing method but is also influenced by the base model architecture.

\section{Additional Tables}
\subsection{Single Edit}

Tables~\ref{tab:instant-results-gen} and~\ref{tab:instant-results-spec} report the complete Compiles@$k$ and Pass@$k$ results for single edits on Generalization and Specificity respectively, across all three models and all values of $k \in \{1, 3, 5\}$. These tables extend Table~\ref{tab:instant-results-gen-spec-at5}, which reports only Pass@5 and Compiles@5, providing the full picture across all $k$ values.

\begin{table*}[h!]
\centering
\scriptsize
\setlength{\tabcolsep}{2.5pt}
\begin{tabular}{l|ccc|ccc|ccc|ccc|ccc|ccc}
\toprule
& \multicolumn{6}{c|}{\textbf{CodeLlama}} & \multicolumn{6}{c|}{\textbf{CodeQwen}} & \multicolumn{6}{c}{\textbf{DeepSeek}} \\
& \multicolumn{3}{c|}{Compiles@k} & \multicolumn{3}{c|}{Pass@k} & \multicolumn{3}{c|}{Compiles@k} & \multicolumn{3}{c|}{Pass@k} & \multicolumn{3}{c|}{Compiles@k} & \multicolumn{3}{c}{Pass@k} \\
\textbf{Method} & 1 & 3 & 5 & 1 & 3 & 5 & 1 & 3 & 5 & 1 & 3 & 5 & 1 & 3 & 5 & 1 & 3 & 5 \\
\midrule
pre-edit & 0.96 & 1.00 & 1.00 & 0.74 & 0.93 & 1.00 & 0.99 & 1.00 & 1.00 & 0.81 & 0.95 & 1.00 & 0.86 & 0.97 & 1.00 & 0.69 & 0.90 & 1.00 \\
\midrule
FT-L & 0.70 & 0.74 & 0.76 & 0.48 & 0.60 & 0.63 & 0.69 & 0.75 & 0.76 & 0.54 & 0.64 & 0.68 & 0.67 & 0.79 & 0.82 & 0.50 & 0.62 & 0.66 \\
GRACE & 0.69 & 0.73 & 0.74 & 0.48 & 0.60 & 0.64 & 0.68 & 0.75 & 0.77 & 0.54 & 0.65 & 0.69 & 0.69 & 0.80 & 0.83 & 0.49 & 0.62 & 0.66 \\
A-GRACE & 0.70 & 0.74 & 0.75 & 0.46 & 0.58 & 0.62 & 0.67 & 0.74 & 0.76 & 0.54 & 0.63 & 0.66 & 0.67 & 0.79 & 0.81 & 0.49 & 0.62 & 0.64 \\
MALMEN & 0.58 & 0.69 & 0.71 & 0.28 & 0.39 & 0.44 & 0.64 & 0.71 & 0.73 & 0.38 & 0.47 & 0.51 & 0.70 & 0.78 & 0.79 & 0.46 & 0.54 & 0.57 \\
MEMIT & 0.38 & 0.46 & 0.50 & 0.17 & 0.23 & 0.26 & 0.65 & 0.71 & 0.73 & 0.46 & 0.55 & 0.58 & 0.28 & 0.39 & 0.45 & 0.16 & 0.21 & 0.23 \\
PMET & 0.50 & 0.59 & 0.62 & 0.22 & 0.29 & 0.32 & 0.65 & 0.71 & 0.74 & 0.51 & 0.61 & 0.65 & 0.34 & 0.44 & 0.49 & 0.18 & 0.23 & 0.25 \\
ROME & 0.50 & 0.61 & 0.66 & 0.24 & 0.32 & 0.35 & 0.67 & 0.73 & 0.74 & 0.43 & 0.52 & 0.55 & 0.30 & 0.39 & 0.42 & 0.17 & 0.23 & 0.26 \\
\bottomrule
\end{tabular}
\caption{Model performance for single edits on Generalization across all $k$ values.}
\label{tab:instant-results-gen}
\end{table*}

\begin{table*}[h!]
\centering
\scriptsize
\setlength{\tabcolsep}{2.5pt}
\begin{tabular}{l|ccc|ccc|ccc|ccc|ccc|ccc}
\toprule
& \multicolumn{6}{c|}{\textbf{CodeLlama}} & \multicolumn{6}{c|}{\textbf{CodeQwen}} & \multicolumn{6}{c}{\textbf{DeepSeek}} \\
& \multicolumn{3}{c|}{Compiles@k} & \multicolumn{3}{c|}{Pass@k} & \multicolumn{3}{c|}{Compiles@k} & \multicolumn{3}{c|}{Pass@k} & \multicolumn{3}{c|}{Compiles@k} & \multicolumn{3}{c}{Pass@k} \\
\textbf{Method} & 1 & 3 & 5 & 1 & 3 & 5 & 1 & 3 & 5 & 1 & 3 & 5 & 1 & 3 & 5 & 1 & 3 & 5 \\
\midrule
pre-edit & 0.96 & 0.99 & 1.00 & 0.71 & 0.92 & 1.00 & 0.98 & 1.00 & 1.00 & 0.80 & 0.95 & 1.00 & 0.80 & 0.95 & 1.00 & 0.61 & 0.86 & 1.00 \\
\midrule
FT-L & 0.96 & 1.00 & 1.00 & 0.67 & 0.83 & 0.88 & 0.98 & 1.00 & 1.00 & 0.76 & 0.88 & 0.92 & 0.79 & 0.93 & 0.96 & 0.58 & 0.73 & 0.80 \\
GRACE & 0.96 & 1.00 & 1.00 & 0.70 & 0.86 & 0.92 & 0.98 & 1.00 & 1.00 & 0.77 & 0.90 & 0.93 & 0.77 & 0.91 & 0.95 & 0.57 & 0.75 & 0.83 \\
A-GRACE & 0.97 & 1.00 & 1.00 & 0.71 & 0.86 & 0.90 & 0.98 & 1.00 & 1.00 & 0.78 & 0.90 & 0.94 & 0.78 & 0.90 & 0.94 & 0.57 & 0.74 & 0.82 \\
MALMEN & 0.86 & 0.94 & 0.95 & 0.49 & 0.63 & 0.68 & 0.84 & 0.89 & 0.91 & 0.53 & 0.63 & 0.67 & 0.76 & 0.87 & 0.89 & 0.43 & 0.56 & 0.61 \\
MEMIT & 0.53 & 0.63 & 0.66 & 0.27 & 0.35 & 0.37 & 0.90 & 0.94 & 0.95 & 0.65 & 0.77 & 0.81 & 0.36 & 0.47 & 0.51 & 0.21 & 0.26 & 0.28 \\
PMET & 0.68 & 0.76 & 0.79 & 0.35 & 0.44 & 0.47 & 0.95 & 0.98 & 0.98 & 0.75 & 0.88 & 0.91 & 0.41 & 0.55 & 0.60 & 0.23 & 0.32 & 0.37 \\
ROME & 0.71 & 0.82 & 0.87 & 0.34 & 0.43 & 0.45 & 0.91 & 0.94 & 0.96 & 0.60 & 0.72 & 0.75 & 0.39 & 0.52 & 0.57 & 0.23 & 0.32 & 0.36 \\
\bottomrule
\end{tabular}
\caption{Model performance for single edits on Specificity across all $k$ values.}
\label{tab:instant-results-spec}
\end{table*}

Across all methods and models, Pass@$k$ and Compiles@$k$ improve consistently as $k$ increases. For FT-L, GRACE, and A-GRACE, both Compiles@$k$ and Pass@$k$ on Specificity remain close to pre-edit levels across all values of $k$, indicating that these methods preserve both compilability and correctness on unrelated problems. In contrast, the remaining methods (MALMEN, MEMIT, PMET, ROME) show significant degradation on Specificity relative to pre-edit, indicating that these methods introduce non-trivial collateral damage to unrelated problems. Finally, degradation is more pronounced on Generalization than Specificity across all methods, confirming that generalizing to new problems sharing the edited API is a harder challenge than preserving performance on unrelated ones.

\subsection{Workarounds}
\begin{table*}[h!]
\centering
\scriptsize
\setlength{\tabcolsep}{2.5pt}
\begin{tabular}{l|ccc|ccc|ccc|ccc|ccc|ccc}
\toprule
& \multicolumn{6}{c|}{\textbf{CodeLlama}} & \multicolumn{6}{c|}{\textbf{CodeQwen}} & \multicolumn{6}{c}{\textbf{DeepSeek}} \\
& \multicolumn{3}{c|}{Compiles@k} & \multicolumn{3}{c|}{Pass@k} & \multicolumn{3}{c|}{Compiles@k} & \multicolumn{3}{c|}{Pass@k} & \multicolumn{3}{c|}{Compiles@k} & \multicolumn{3}{c}{Pass@k} \\
\textbf{Method} & 1 & 3 & 5 & 1 & 3 & 5 & 1 & 3 & 5 & 1 & 3 & 5 & 1 & 3 & 5 & 1 & 3 & 5 \\
\midrule
pre-edit & 0.96 & 1.00 & 1.00 & 0.78 & 0.94 & 1.00 & 0.99 & 1.00 & 1.00 & 0.82 & 0.96 & 1.00 & 0.85 & 0.97 & 1.00 & 0.68 & 0.90 & 1.00 \\
\midrule
FT-L & 0.43 & 0.49 & 0.51 & 0.30 & 0.38 & 0.40 & 0.45 & 0.54 & 0.57 & 0.33 & 0.43 & 0.46 & 0.40 & 0.56 & 0.60 & 0.27 & 0.36 & 0.38 \\
GRACE & 0.42 & 0.46 & 0.49 & 0.30 & 0.39 & 0.41 & 0.44 & 0.54 & 0.58 & 0.33 & 0.43 & 0.48 & 0.46 & 0.59 & 0.64 & 0.28 & 0.35 & 0.38 \\
A-GRACE & 0.43 & 0.48 & 0.50 & 0.30 & 0.40 & 0.43 & 0.43 & 0.53 & 0.57 & 0.33 & 0.40 & 0.44 & 0.44 & 0.57 & 0.60 & 0.25 & 0.33 & 0.34 \\
MALMEN & 0.38 & 0.48 & 0.51 & 0.18 & 0.25 & 0.28 & 0.50 & 0.58 & 0.59 & 0.24 & 0.32 & 0.35 & 0.48 & 0.58 & 0.60 & 0.29 & 0.33 & 0.36 \\
MEMIT & 0.23 & 0.31 & 0.34 & 0.10 & 0.14 & 0.16 & 0.46 & 0.55 & 0.58 & 0.29 & 0.36 & 0.38 & 0.23 & 0.36 & 0.43 & 0.11 & 0.16 & 0.19 \\
PMET & 0.38 & 0.46 & 0.49 & 0.12 & 0.18 & 0.20 & 0.41 & 0.51 & 0.55 & 0.32 & 0.40 & 0.44 & 0.26 & 0.38 & 0.45 & 0.11 & 0.15 & 0.17 \\
ROME & 0.33 & 0.46 & 0.54 & 0.16 & 0.23 & 0.26 & 0.52 & 0.60 & 0.62 & 0.26 & 0.34 & 0.37 & 0.26 & 0.37 & 0.42 & 0.12 & 0.16 & 0.19 \\
\bottomrule
\end{tabular}
\caption{Model performance for single edits on the canonically filtered Generalization set across all k values.}
\label{tab:canonical-instant-gen}
\end{table*}

\begin{table*}[h!]
\centering
\scriptsize
\setlength{\tabcolsep}{2.5pt}
\begin{tabular}{l|ccc|ccc|ccc|ccc|ccc|ccc}
\toprule
& \multicolumn{6}{c|}{\textbf{CodeLlama}} & \multicolumn{6}{c|}{\textbf{CodeQwen}} & \multicolumn{6}{c}{\textbf{DeepSeek}} \\
& \multicolumn{3}{c|}{Adoption@k} & \multicolumn{3}{c|}{Workaround@k} & \multicolumn{3}{c|}{Adoption@k} & \multicolumn{3}{c|}{Workaround@k} & \multicolumn{3}{c|}{Adoption@k} & \multicolumn{3}{c}{Workaround@k} \\
\textbf{Method} & @1 & @3 & @5 & @1 & @3 & @5 & @1 & @3 & @5 & @1 & @3 & @5 & @1 & @3 & @5 & @1 & @3 & @5 \\
\midrule
FT-L & 0.11 & 0.15 & 0.17 & 0.19 & 0.23 & 0.23 & 0.14 & 0.16 & 0.18 & 0.20 & 0.27 & 0.29 & 0.12 & 0.15 & 0.17 & 0.15 & 0.20 & 0.21 \\
GRACE & 0.10 & 0.12 & 0.13 & 0.20 & 0.26 & 0.28 & 0.14 & 0.16 & 0.17 & 0.19 & 0.27 & 0.31 & 0.13 & 0.17 & 0.19 & 0.14 & 0.19 & 0.19 \\
A-GRACE & 0.12 & 0.16 & 0.17 & 0.19 & 0.24 & 0.26 & 0.13 & 0.16 & 0.16 & 0.20 & 0.25 & 0.27 & 0.11 & 0.14 & 0.15 & 0.14 & 0.18 & 0.19 \\
MALMEN & 0.08 & 0.10 & 0.11 & 0.10 & 0.15 & 0.17 & 0.07 & 0.08 & 0.10 & 0.17 & 0.23 & 0.25 & 0.11 & 0.15 & 0.17 & 0.17 & 0.19 & 0.19 \\
MEMIT & 0.07 & 0.09 & 0.10 & 0.03 & 0.05 & 0.06 & 0.12 & 0.14 & 0.15 & 0.17 & 0.22 & 0.23 & 0.08 & 0.10 & 0.11 & 0.03 & 0.06 & 0.08 \\
PMET & 0.06 & 0.09 & 0.10 & 0.06 & 0.09 & 0.10 & 0.13 & 0.15 & 0.16 & 0.19 & 0.25 & 0.29 & 0.09 & 0.12 & 0.13 & 0.02 & 0.03 & 0.04 \\
ROME & 0.06 & 0.08 & 0.09 & 0.10 & 0.15 & 0.17 & 0.10 & 0.12 & 0.13 & 0.16 & 0.22 & 0.24 & 0.07 & 0.10 & 0.11 & 0.05 & 0.07 & 0.08 \\
\bottomrule
\end{tabular}
\caption{Successful Adoption and Workaround after a single edit on the canonically filtered Generalization set across k values.}
\label{tab:migration_metrics_instant}
\end{table*}

Tables~\ref{tab:canonical-instant-gen} and~\ref{tab:migration_metrics_instant} extend Table~\ref{tab:canonical-instant-gen-at5} to all $k \in {1, 3, 5}$, reporting Compiles@$k$ and Pass@$k$ on the canonical Generalization dataset and their decomposition into Adoption@$k$ and Workaround@$k$ respectively, distinguishing between solutions that correctly adopt the new API and those that pass by avoiding it.
Across all methods and models, Workaround@$k$ is consistently higher than Adoption@$k$, indicating that models more frequently find alternative implementations that avoid the edited API than correctly adopt it. The gap between Workaround@$k$ and Adoption@$k$ is most pronounced for FT-L, GRACE, and A-GRACE, while weight-modifying methods (MEMIT, PMET, ROME) show low values on both metrics, suggesting that these methods neither adopt the new API nor successfully find workarounds.

\subsection{Non-workaroundable APIs}
\begin{table*}[h!]
\centering
\scriptsize
\setlength{\tabcolsep}{2.5pt}
\begin{tabular}{l|ccc|ccc|ccc|ccc|ccc|ccc}
\toprule
& \multicolumn{6}{c|}{\textbf{CodeLlama}} & \multicolumn{6}{c|}{\textbf{CodeQwen}} & \multicolumn{6}{c}{\textbf{DeepSeek}} \\
& \multicolumn{3}{c|}{Compiles@k} & \multicolumn{3}{c|}{Pass@k} & \multicolumn{3}{c|}{Compiles@k} & \multicolumn{3}{c|}{Pass@k} & \multicolumn{3}{c|}{Compiles@k} & \multicolumn{3}{c}{Pass@k} \\
\textbf{Method} & 1 & 3 & 5 & 1 & 3 & 5 & 1 & 3 & 5 & 1 & 3 & 5 & 1 & 3 & 5 & 1 & 3 & 5 \\
\midrule
pre-edit & 1.00 & 1.00 & 1.00 & 0.40 & 0.40 & 0.40 & 1.00 & 1.00 & 1.00 & 0.54 & 0.65 & 0.70 & 1.00 & 1.00 & 1.00 & 0.66 & 0.75 & 0.80 \\
\midrule
FT-L & 0.22 & 0.26 & 0.30 & 0.18 & 0.20 & 0.20 & 0.32 & 0.40 & 0.40 & 0.12 & 0.16 & 0.20 & 0.28 & 0.30 & 0.30 & 0.20 & 0.20 & 0.20 \\
GRACE & 0.20 & 0.20 & 0.20 & 0.18 & 0.20 & 0.20 & 0.28 & 0.30 & 0.30 & 0.10 & 0.10 & 0.10 & 0.30 & 0.36 & 0.40 & 0.20 & 0.20 & 0.20 \\
A-GRACE & 0.20 & 0.20 & 0.20 & 0.20 & 0.20 & 0.20 & 0.20 & 0.20 & 0.20 & 0.10 & 0.10 & 0.10 & 0.26 & 0.35 & 0.40 & 0.20 & 0.20 & 0.20 \\
MALMEN & 0.20 & 0.20 & 0.20 & 0.20 & 0.20 & 0.20 & 0.52 & 0.65 & 0.70 & 0.00 & 0.00 & 0.00 & 0.28 & 0.36 & 0.40 & 0.20 & 0.20 & 0.20 \\
MEMIT & 0.00 & 0.00 & 0.00 & 0.00 & 0.00 & 0.00 & 0.22 & 0.32 & 0.40 & 0.00 & 0.00 & 0.00 & 0.00 & 0.00 & 0.00 & 0.00 & 0.00 & 0.00 \\
PMET & 0.18 & 0.26 & 0.30 & 0.02 & 0.06 & 0.10 & 0.24 & 0.32 & 0.40 & 0.10 & 0.10 & 0.10 & 0.16 & 0.37 & 0.50 & 0.00 & 0.00 & 0.00 \\
ROME & 0.32 & 0.47 & 0.50 & 0.08 & 0.18 & 0.20 & 0.20 & 0.20 & 0.20 & 0.18 & 0.20 & 0.20 & 0.08 & 0.24 & 0.40 & 0.02 & 0.06 & 0.10 \\
\bottomrule
\end{tabular}
\caption{Performance for single edits on the non-workaroundable subset across all k values. Only solutions using the target API can succeed.}
\label{tab:non-workaroundable-apis}
\end{table*}

Table~\ref{tab:non-workaroundable-apis} expands Table~\ref{tab:non-workaroundable-apis_at5} to all values of $k \in \{1, 3, 5\}$, reporting Compiles@$k$ and Pass@$k$ on the non-workaroundable dataset. The most notable pattern is the large gap between Compiles@$k$ and Pass@$k$ across almost all methods and models. Importantly, high Compiles@$k$ here does not indicate successful API adoption. Rather, it reflects that models produce syntactically valid code that did not adopt the new API, whether by generating code that does not serve the intended purpose, using the wrong function signature or via workaround exception, as illustrated by the examples in Appendix~\ref{app:non-workaround-examples}. This is illustrated by MALMEN on CodeQwen, where Compiles@$k$ reaches 0.70 at $k=5$ while Pass@$k$ remains 0.00. MEMIT fully collapses on CodeLlama and DeepSeek with both metrics at 0.00, while all other methods show at least some compilable outputs, though with consistently low Pass@$k$ values relative to pre-edit.

\clearpage

\subsection{Successive Edits}
\begin{table*}[h!]
\centering
\scriptsize
\setlength{\tabcolsep}{2.5pt}
\begin{tabular}{l|ccc|ccc|ccc|ccc|ccc|ccc}
\toprule
& \multicolumn{6}{c|}{\textbf{CodeLlama}} & \multicolumn{6}{c|}{\textbf{CodeQwen}} & \multicolumn{6}{c}{\textbf{DeepSeek}} \\
& \multicolumn{3}{c|}{Compiles@k} & \multicolumn{3}{c|}{Pass@k} & \multicolumn{3}{c|}{Compiles@k} & \multicolumn{3}{c|}{Pass@k} & \multicolumn{3}{c|}{Compiles@k} & \multicolumn{3}{c}{Pass@k} \\
\textbf{Method} & 1 & 3 & 5 & 1 & 3 & 5 & 1 & 3 & 5 & 1 & 3 & 5 & 1 & 3 & 5 & 1 & 3 & 5 \\
\midrule
pre-edit & 0.96 & 1.00 & 1.00 & 0.74 & 0.93 & 1.00 & 0.99 & 1.00 & 1.00 & 0.81 & 0.95 & 1.00 & 0.86 & 0.97 & 1.00 & 0.69 & 0.90 & 1.00 \\
\midrule
FT-L & 0.66 & 0.73 & 0.76 & 0.19 & 0.23 & 0.26 & 0.75 & 0.86 & 0.88 & 0.31 & 0.39 & 0.42 & 0.30 & 0.47 & 0.54 & 0.16 & 0.24 & 0.29 \\
GRACE & 0.59 & 0.62 & 0.63 & 0.40 & 0.49 & 0.52 & 0.61 & 0.66 & 0.67 & 0.48 & 0.55 & 0.58 & 0.55 & 0.67 & 0.70 & 0.39 & 0.51 & 0.55 \\
A-GRACE & 0.02 & 0.03 & 0.04 & 0.01 & 0.02 & 0.02 & 0.61 & 0.68 & 0.72 & 0.49 & 0.58 & 0.61 & 0.02 & 0.02 & 0.02 & 0.02 & 0.02 & 0.02 \\
MALMEN & 0.20 & 0.45 & 0.61 & 0.00 & 0.00 & 0.00 & 0.00 & 0.00 & 0.00 & 0.00 & 0.00 & 0.00 & 0.00 & 0.00 & 0.00 & 0.00 & 0.00 & 0.00 \\
MEMIT & 0.00 & 0.00 & 0.00 & 0.00 & 0.00 & 0.00 & 0.00 & 0.00 & 0.00 & 0.00 & 0.00 & 0.00 & 0.00 & 0.00 & 0.00 & 0.00 & 0.00 & 0.00 \\
PMET & 0.00 & 0.00 & 0.00 & 0.00 & 0.00 & 0.00 & 1.00 & 1.00 & 1.00 & 0.00 & 0.00 & 0.00 & 0.00 & 0.00 & 0.00 & 0.00 & 0.00 & 0.00 \\
ROME & 0.11 & 0.16 & 0.18 & 0.00 & 0.00 & 0.00 & 0.00 & 0.00 & 0.00 & 0.00 & 0.00 & 0.00 & 0.00 & 0.00 & 0.00 & 0.00 & 0.00 & 0.00 \\
\bottomrule
\end{tabular}
\caption{Model performance for successive edits on Generalization across all k values.}
\label{tab:sequential-results-gen}
\end{table*}

\begin{table*}[h!]
\centering
\scriptsize
\setlength{\tabcolsep}{2.5pt}
\begin{tabular}{l|ccc|ccc|ccc|ccc|ccc|ccc}
\toprule
& \multicolumn{6}{c|}{\textbf{CodeLlama}} & \multicolumn{6}{c|}{\textbf{CodeQwen}} & \multicolumn{6}{c}{\textbf{DeepSeek}} \\
& \multicolumn{3}{c|}{Compiles@k} & \multicolumn{3}{c|}{Pass@k} & \multicolumn{3}{c|}{Compiles@k} & \multicolumn{3}{c|}{Pass@k} & \multicolumn{3}{c|}{Compiles@k} & \multicolumn{3}{c}{Pass@k} \\
\textbf{Method} & 1 & 3 & 5 & 1 & 3 & 5 & 1 & 3 & 5 & 1 & 3 & 5 & 1 & 3 & 5 & 1 & 3 & 5 \\
\midrule
pre-edit & 0.96 & 0.99 & 1.00 & 0.71 & 0.92 & 1.00 & 0.98 & 1.00 & 1.00 & 0.80 & 0.95 & 1.00 & 0.80 & 0.95 & 1.00 & 0.61 & 0.86 & 1.00 \\
\midrule
FT-L & 0.86 & 0.92 & 0.93 & 0.29 & 0.37 & 0.40 & 0.93 & 0.99 & 0.99 & 0.23 & 0.32 & 0.35 & 0.33 & 0.49 & 0.57 & 0.18 & 0.28 & 0.33 \\
GRACE & 0.96 & 1.00 & 1.00 & 0.70 & 0.85 & 0.92 & 0.98 & 1.00 & 1.00 & 0.78 & 0.91 & 0.95 & 0.79 & 0.93 & 0.97 & 0.57 & 0.73 & 0.81 \\
A-GRACE & 0.00 & 0.00 & 0.00 & 0.00 & 0.00 & 0.00 & 0.96 & 1.00 & 1.00 & 0.75 & 0.89 & 0.95 & 0.00 & 0.00 & 0.00 & 0.00 & 0.00 & 0.00 \\
MALMEN & 0.15 & 0.38 & 0.53 & 0.00 & 0.00 & 0.00 & 0.00 & 0.00 & 0.00 & 0.00 & 0.00 & 0.00 & 0.00 & 0.00 & 0.00 & 0.00 & 0.00 & 0.00 \\
MEMIT & 0.00 & 0.00 & 0.00 & 0.00 & 0.00 & 0.00 & 0.00 & 0.00 & 0.00 & 0.00 & 0.00 & 0.00 & 0.00 & 0.00 & 0.00 & 0.00 & 0.00 & 0.00 \\
PMET & 0.00 & 0.00 & 0.00 & 0.00 & 0.00 & 0.00 & 0.00 & 1.00 & 1.00 & 0.00 & 0.00 & 0.00 & 0.00 & 0.00 & 0.00 & 0.00 & 0.00 & 0.00 \\
ROME & 0.07 & 0.12 & 0.15 & 0.00 & 0.00 & 0.00 & 0.00 & 0.00 & 0.00 & 0.00 & 0.00 & 0.00 & 0.00 & 0.00 & 0.00 & 0.00 & 0.00 & 0.00 \\
\bottomrule
\end{tabular}
\caption{Model performance for successive edits on Specificity across all k values.}
\label{tab:sequential-results-spec}
\end{table*}

Tables~\ref{tab:sequential-results-gen} and~\ref{tab:sequential-results-spec} expand Table~\ref{tab:seq-results-gen-spec-agg} to all values of $k \in {1, 3, 5}$, reporting Compiles@$k$ and Pass@$k$ for successive edits on Generalization and Specificity respectively. GRACE is the only method to maintain meaningful Pass@$k$ across all models on both splits, preserving near pre-edit Compiles@$k$ and Pass@$k$ on Specificity in particular. FT-L partially survives on both splits but with substantially lower Pass@$k$ than pre-edit across all models. A-GRACE collapses completely for CodeLlama and DeepSeek on both splits while surviving for CodeQwen on both splits, highlighting its instability across models. MALMEN and ROME produce some compilable outputs only for CodeLlama on both splits, with Pass@$k$ = 0.00 across all models and both splits, indicating that even where compilable outputs are generated they all fail at runtime. For CodeQwen and DeepSeek, both methods produce no compilable outputs at all. MEMIT collapses entirely with 0.00 on all metrics across all models and both splits. Finally, PMET on CodeQwen shows the same anomaly noted in the Shapley analysis: Compiles@$k$ reaches 1.00 on Generalization yet Pass@$k$ remains 0.00, confirming that all generated outputs are compilable but fail at runtime.

\section{Code Samples}
\label{appendix:samples}
\subsection{Compilation Errors} \label{app:code-s2_error}

\noindent\begin{tcolorbox}[
    fonttitle=\bfseries\small,
    colframe=black!40!white,
    colback=black!5!white,
    coltitle=black,
    boxrule=0.5pt, arc=2pt,
    left=4pt, right=4pt, top=2pt, bottom=2pt,
    title=Task Context]
\textbf{Prompt:} Implement \texttt{evil(n: int) -> str}, which determines if a number is Evil (even number of 1s in binary) or Odious (odd number of 1s in binary representation).\\[4pt]
\textbf{S2 Modification:} \texttt{bin(n)} $\rightarrow$ \texttt{bin(n, prefix\_required=True)}, where a new required parameter \texttt{prefix\_required} is added, making the old call signature invalid.
\end{tcolorbox}
\vspace{4pt}

\noindent\begin{tcolorbox}[
    fonttitle=\bfseries\small,
    colframe=preEditColor!70!black,
    colback=preEditColor!20!white,
    coltitle=black,
    boxrule=0.5pt, arc=2pt,
    left=4pt, right=4pt, top=2pt, bottom=2pt,
    title=Pre-edit (Reference) | CodeLlama]
\begin{Verbatim}[commandchars=\\\{\},fontsize=\footnotesize]
def evil(n: int) -> str:
    binary_rep = \targetapi{bin(n)}
    num_ones = binary_rep.count('1')
    if num_ones % 2 == 0:
        return "It's Evil!"
    else:
        return "It's Odious!"
\end{Verbatim}
\end{tcolorbox}
\vspace{-8pt}
\captionof{figure}{Correct reference implementation prior to API editing.
\targetapi{\texttt{bin(n)}} marks the old API call that must be updated to
\texttt{bin(n, prefix\_required=True)}.}
\vspace{16pt}

\noindent\begin{tcolorbox}[
    fonttitle=\bfseries\small,
    colframe=preEditColor!70!black,
    colback=preEditColor!20!white,
    coltitle=black,
    boxrule=0.5pt, arc=2pt,
    left=4pt, right=4pt, top=2pt, bottom=2pt,
    title=FT-L / GRACE / A-GRACE / MALMEN  | CodeLlama | S2-Error]
\begin{Verbatim}[commandchars=\\\{\},fontsize=\footnotesize]
def evil(n: int) -> str:
    binary_representation = \oldapi{bin(n)}
    num_ones = binary_representation.count('1')
    if num_ones % 2 == 0:
        return "It's Evil!"
    else:
        return "It's Odious!"
\end{Verbatim}
\end{tcolorbox}
\vspace{-8pt}
\captionof{figure}{FT-L, GRACE, A-GRACE, and MALMEN all fail to adopt the new API.
The generated code is syntactically valid but \oldapi{\texttt{bin(n)}} remains
unchanged, resulting in a failed API adoption: \texttt{bin()} missing
required keyword argument \texttt{prefix\_required}.}
\vspace{16pt}

\noindent\begin{tcolorbox}[
    fonttitle=\bfseries\small,
    colframe=preEditColor!70!black,
    colback=preEditColor!20!white,
    coltitle=black,
    boxrule=0.5pt, arc=2pt,
    left=4pt, right=4pt, top=2pt, bottom=2pt,
    title=MEMIT | CodeLlama | SyntaxError]
\begin{Verbatim}[commandchars=\\\{\},fontsize=\footnotesize]
\oldapi{Odious(n):}
\oldapi{    ):}
    return "It's Evil!" if n(n, 1) else "It's Odious!"
    return "It's Evil!" if n(n, 1) else "It's Odious!"
\end{Verbatim}
\end{tcolorbox}
\vspace{-8pt}
\captionof{figure}{MEMIT produces structurally corrupted output. Line 1 is missing \texttt{def} and has an invalid signature. Line 2 contains a stray closing parenthesis. Both highlighted tokens trigger a \texttt{SyntaxError} at compile time before any execution occurs. The duplicated return statements are unreachable.}
\vspace{16pt}

\noindent\begin{tcolorbox}[
    fonttitle=\bfseries\small,
    colframe=preEditColor!70!black,
    colback=preEditColor!20!white,
    coltitle=black,
    boxrule=0.5pt, arc=2pt,
    left=4pt, right=4pt, top=2pt, bottom=2pt,
    title=ROME | CodeLlama | IndentationError]
\begin{Verbatim}[commandchars=\\\{\},fontsize=\footnotesize]
def evil(n: int) -> str:
\oldapi{    Empty body, generation collapsed entirely}
\end{Verbatim}
\end{tcolorbox}
\vspace{-8pt}
\captionof{figure}{ROME exhibits complete generation collapse: the function signature
is intact but the body is empty, and since no valid indented statement follows the
\texttt{def} block, Python raises an \texttt{IndentationError} at compile time.}
\vspace{16pt}

\noindent\begin{tcolorbox}[
    fonttitle=\bfseries\small,
    colframe=preEditColor!70!black,
    colback=preEditColor!20!white,
    coltitle=black,
    boxrule=0.5pt, arc=2pt,
    left=4pt, right=4pt, top=2pt, bottom=2pt,
    title=PMET | CodeLlama | SyntaxError]
\begin{Verbatim}[commandchars=\\\{\},fontsize=\footnotesize]
def evil(n):
    return (
\oldapi{        f[n]}
\oldapi{        if n else}
\oldapi{        f[n]}
\oldapi{        if n else}
\oldapi{        f[n]}
\oldapi{        ...  pattern repeats 12 times, truncated}
\end{Verbatim}
\end{tcolorbox}
\vspace{-8pt}
\captionof{figure}{PMET degrades into a repetitive ternary chain referencing
undefined variable \texttt{f}. The \texttt{return (} on line 2 is never closed,
causing a \texttt{SyntaxError} at compile time.}
\vspace{16pt}
\clearpage
\subsection{Workarounds}
\label{app:workaround-examples}

\noindent\begin{tcolorbox}[
    fonttitle=\bfseries\small,
    colframe=black!40!white,
    colback=black!5!white,
    coltitle=black,
    boxrule=0.5pt, arc=2pt,
    left=4pt, right=4pt, top=2pt, bottom=2pt,
    title=Task Context]
\textbf{Prompt:} Implement \texttt{prime\_or\_composite(n: int) -> str}, which checks if a number is prime (return \texttt{"Probable Prime"}) or composite (return \texttt{"Composite"}) for $N \leq 10^{100}$. An efficient algorithm is required given the large upper bound.\\[4pt]
\textbf{S4 Modification:} \texttt{random.randint(a, b)} $\rightarrow$ \texttt{random.randint(a, b)[0]}, where the function now returns a tuple and the result must be indexed at \texttt{[0]} to obtain the integer value.
\end{tcolorbox}
\vspace{4pt}

\noindent\begin{tcolorbox}[
    fonttitle=\bfseries\small,
    colframe=preEditColor!70!black,
    colback=preEditColor!20!white,
    coltitle=black,
    boxrule=0.5pt, arc=2pt,
    left=4pt, right=4pt, top=2pt, bottom=2pt,
    title=Pre-edit (Reference) | CodeQwen]
\begin{Verbatim}[commandchars=\\\{\},fontsize=\footnotesize]
import random

def prime_or_composite(n: int) -> str:
    if n < 2:
        return "Composite"
    if n == 2:
        return "Probable Prime"
    if n % 2 == 0:
        return "Composite"
    r, d = 0, n - 1
    while d % 2 == 0:
        r += 1
        d //= 2
    for _ in range(5):
        a = \targetapi{random.randint(2, n - 1)}
        x = pow(a, d, n)
        if x == 1 or x == n - 1:
            continue
        for _ in range(r - 1):
            x = pow(x, 2, n)
            if x == n - 1:
                break
        else:
            return "Composite"
    return "Probable Prime"
\end{Verbatim}
\end{tcolorbox}
\vspace{-8pt}
\captionof{figure}{Correct reference implementation using the Miller-Rabin primality test, where
\targetapi{\texttt{random.randint(2, n-1)}} samples a random witness per round and must be
updated to \targetapi{\texttt{random.randint(2, n-1)[0]}} after the S4 modification.}
\vspace{16pt}

\noindent\begin{tcolorbox}[
    fonttitle=\bfseries\small,
    colframe=preEditColor!70!black,
    colback=preEditColor!20!white,
    coltitle=black,
    boxrule=0.5pt, arc=2pt,
    left=4pt, right=4pt, top=2pt, bottom=2pt,
    title=A-GRACE | CodeQwen | Workaround]
\begin{Verbatim}[commandchars=\\\{\},fontsize=\footnotesize]
import random

def prime_or_composite(n: int) -> str:
    if n <= 1:
        return "Composite"
    if n <= 3:
        return "Probable Prime"
    if n % 2 == 0 or n % 3 == 0:
        return "Composite"
    i = 5
    while i * i <= n:
        if n % i == 0 or n % (i + 2) == 0:
            return "Composite"
        i += 6
    return "Probable Prime"
\end{Verbatim}
\end{tcolorbox}
\vspace{-8pt}
\captionof{figure}{A-GRACE avoids the modified API by substituting trial division (checking
divisors of the form $6k \pm 1$ up to $\sqrt{n}$) in place of Miller-Rabin, eliminating
any need for \targetapi{\texttt{random.randint}}. The solution is algorithmically correct but
infeasible for $N \leq 10^{100}$, passing only due to small test inputs.}
\vspace{16pt}

\noindent\begin{tcolorbox}[
    fonttitle=\bfseries\small,
    colframe=black!40!white,
    colback=black!5!white,
    coltitle=black,
    boxrule=0.5pt, arc=2pt,
    left=4pt, right=4pt, top=2pt, bottom=2pt,
    title=Task Context]
\textbf{Prompt:} Implement \texttt{sum\_of\_odd\_factors(n: int) -> int}, which returns the sum of all odd factors of a given integer $n$.\\[4pt]
\textbf{S4 Modification:} \texttt{math.sqrt(n)} $\rightarrow$ \texttt{math.sqrt(n)[0]}, where the function now returns a tuple and the result must be indexed at \texttt{[0]} to obtain the numeric value.
\end{tcolorbox}
\vspace{4pt}

\noindent\begin{tcolorbox}[
    fonttitle=\bfseries\small,
    colframe=preEditColor!70!black,
    colback=preEditColor!20!white,
    coltitle=black,
    boxrule=0.5pt, arc=2pt,
    left=4pt, right=4pt, top=2pt, bottom=2pt,
    title=Pre-edit (Reference) | CodeQwen]
\begin{Verbatim}[commandchars=\\\{\},fontsize=\footnotesize]
import math

def sum_of_odd_factors(n: int) -> int:
    result = 0
    for i in range(1, int(\targetapi{math.sqrt(n)}) + 1):
        if n % i == 0:
            if i % 2 != 0:
                result += i
            if (n // i != i) and ((n // i) % 2 != 0):
                result += n // i
    return result
\end{Verbatim}
\end{tcolorbox}
\vspace{-8pt}
\captionof{figure}{Correct reference implementation iterating up to $\lfloor\sqrt{n}\rfloor$ via
\targetapi{\texttt{math.sqrt(n)}}, collecting odd factors in $O(\sqrt{n})$ time.
After the S4 modification, this call must be updated to \targetapi{\texttt{math.sqrt(n)[0]}}.}
\vspace{16pt}

\noindent\begin{tcolorbox}[
    fonttitle=\bfseries\small,
    colframe=preEditColor!70!black,
    colback=preEditColor!20!white,
    coltitle=black,
    boxrule=0.5pt, arc=2pt,
    left=4pt, right=4pt, top=2pt, bottom=2pt,
    title=PMET | CodeQwen | Workaround]
\begin{Verbatim}[commandchars=\\\{\},fontsize=\footnotesize]
def sum_of_odd_factors(n: int) -> int:
    sum_factors = 0
    for i in range(1, n+1):
        if n % i == 0 and i % 2 != 0:
            sum_factors += i
    return sum_factors
\end{Verbatim}
\end{tcolorbox}
\vspace{-8pt}
\captionof{figure}{PMET avoids the modified API by replacing the $O(\sqrt{n})$
factor search with a brute-force $O(n)$ loop over all integers up to $n$, eliminating
any need for \targetapi{\texttt{math.sqrt(n)[0]}}. The solution is correct
but does not scale; it passes only due to small test inputs.}
\vspace{16pt}

\clearpage
\subsection{Non-workaroundable APIs}
\label{app:non-workaround-examples}
\vspace{8pt}

\noindent\begin{tcolorbox}[
    fonttitle=\bfseries\small,
    colframe=black!40!white,
    colback=black!5!white,
    coltitle=black,
    boxrule=0.5pt, arc=2pt,
    left=4pt, right=4pt, top=2pt, bottom=2pt,
    title=Task Context]
\textbf{Prompt:} Implement \texttt{count\_gpus\_minus10() -> int}, which counts available CUDA devices using \texttt{torch} and subtracts 10.\\[4pt]
\textbf{S1 Modification:} \texttt{torch.cuda.device\_count()} now accepts an optional argument; the original call signature remains valid.
\end{tcolorbox}
\vspace{4pt}
\noindent\begin{tcolorbox}[
    fonttitle=\bfseries\small,
    colframe=black!40!white,
    colback=black!5!white,
    coltitle=black,
    boxrule=0.5pt, arc=2pt,
    left=4pt, right=4pt, top=2pt, bottom=2pt,
    title=MALMEN | CodeQwen | Wrong Function Name]
\begin{Verbatim}[commandchars=\\\{\},fontsize=\footnotesize]
def \oldapi{count\_gpus\_minus\_10}() -> int:
    return max(0, torch.cuda.device_count() - 10)
\end{Verbatim}
\end{tcolorbox}
\vspace{-8pt}
\captionof{figure}{MALMEN defines \oldapi{\texttt{count\_gpus\_minus\_10}} instead of the required \targetapi{\texttt{count\_gpus\_minus10}}. The code is logically correct and compiles, but raises a \texttt{NameError} at test execution when the expected entry point is called.}
\vspace{16pt}

\noindent\begin{tcolorbox}[
    fonttitle=\bfseries\small,
    colframe=black!40!white,
    colback=black!5!white,
    coltitle=black,
    boxrule=0.5pt, arc=2pt,
    left=4pt, right=4pt, top=2pt, bottom=2pt,
    title=Task Context]
\textbf{Prompt:} Implement \texttt{get\_recursion\_limit\_str() -> str}, which returns the maximum recursion limit as a string using the \texttt{sys} module.\\[4pt]
\textbf{S4 Modification:} \texttt{sys.getrecursionlimit()} $\rightarrow$ \texttt{sys.getrecursionlimit()[0]}, where the function now returns a tuple and the result must be indexed at \texttt{[0]}.
\end{tcolorbox}
\vspace{4pt}
\noindent\begin{tcolorbox}[
    fonttitle=\bfseries\small,
    colframe=black!40!white,
    colback=black!5!white,
    coltitle=black,
    boxrule=0.5pt, arc=2pt,
    left=4pt, right=4pt, top=2pt, bottom=2pt,
    title=GRACE | CodeLlama | Return/Type Mismatch]
\begin{Verbatim}[commandchars=\\\{\},fontsize=\footnotesize]
import sys
def get_recursion_limit_str():
    return str(\oldapi{sys.getrecursionlimit()})
\end{Verbatim}
\end{tcolorbox}
\vspace{-8pt}
\captionof{figure}{GRACE retains the old \oldapi{\texttt{sys.getrecursionlimit()}} call without the required \texttt{[0]} index and raises a Return/Type Mismatch Error.}
\vspace{16pt}

\noindent\begin{tcolorbox}[
    fonttitle=\bfseries\small,
    colframe=black!40!white,
    colback=black!5!white,
    coltitle=black,
    boxrule=0.5pt, arc=2pt,
    left=4pt, right=4pt, top=2pt, bottom=2pt,
    title=Task Context]
\textbf{Prompt:} Implement \texttt{move\_and\_verify(src\_folder, dst\_folder, filename) -> bool}, which moves a file from source to destination and returns \texttt{True} if the source no longer contains the file, using \texttt{shutil}.\\[4pt]
\textbf{S3 Modification:} \texttt{shutil.move(src, dst)} $\rightarrow$ \texttt{shutil.move(dst, src)}, where the argument order is swapped.
\end{tcolorbox}
\vspace{4pt}

\noindent\begin{tcolorbox}[
    fonttitle=\bfseries\small,
    colframe=black!40!white,
    colback=black!5!white,
    coltitle=black,
    boxrule=0.5pt, arc=2pt,
    left=4pt, right=4pt, top=2pt, bottom=2pt,
    title=MALMEN | DeepSeek | Test Failure]
\begin{Verbatim}[commandchars=\\\{\},fontsize=\footnotesize]
def move_and_verify(src_folder: str,
                    dst_folder: str,
                    filename: str) -> bool:
    src = os.path.join(src_folder, filename)
    dst = os.path.join(dst_folder, filename)
\end{Verbatim}
\end{tcolorbox}
\vspace{-8pt}
\captionof{figure}{MALMEN generates syntactically valid code that invokes neither \targetapi{\texttt{shutil.move()}} nor its edited variant, instead only constructing path strings without performing any file operation. The function returns \texttt{None} implicitly, failing all tests with an \texttt{AssertionError}.}
\vspace{16pt}

\clearpage

\subsection{Failure after Adoption}
\vspace{4pt}

\noindent\begin{tcolorbox}[
    fonttitle=\bfseries\small,
    colframe=black!40!white,
    colback=black!5!white,
    coltitle=black,
    boxrule=0.5pt, arc=2pt,
    left=4pt, right=4pt, top=2pt, bottom=2pt,
    title=Task Context]
\textbf{Prompt:} Implement \texttt{superPow(a: int, b: list[int]) -> int}, which computes $a^b \mod 1337$ where $b$ is given as a digit array.\\[4pt]
\textbf{S4 Modification:} \texttt{pow(a, b, mod)} $\rightarrow$ \texttt{pow(a, b, mod)[0]}, where the function now returns a tuple and the result must be indexed at \texttt{[0]}.
\end{tcolorbox}
\vspace{4pt}

\noindent\begin{tcolorbox}[
    fonttitle=\bfseries\small,
    colframe=black!40!white,
    colback=black!5!white,
    coltitle=black,
    boxrule=0.5pt, arc=2pt,
    left=4pt, right=4pt, top=2pt, bottom=2pt,
    title=Pre-edit (Reference) | CodeQwen]
\begin{Verbatim}[commandchars=\\\{\},fontsize=\footnotesize]
def superPow(a: int, b: list[int]) -> int:
    return \targetapi{pow}(a, int("".join(map(str, b))), 1337)
\end{Verbatim}
\end{tcolorbox}
\vspace{-8pt}
\captionof{figure}{Correct reference implementation reconstructing the exponent from the digit
array via \targetapi{\texttt{pow(a, b, mod)}}, which must be updated to
\targetapi{\texttt{pow(a, b, mod)[0]}} after the S4 modification.}
\vspace{16pt}

\noindent\begin{tcolorbox}[
    fonttitle=\bfseries\small,
    colframe=black!40!white,
    colback=black!5!white,
    coltitle=black,
    boxrule=0.5pt, arc=2pt,
    left=4pt, right=4pt, top=2pt, bottom=2pt,
    title=MALMEN | CodeQwen | Test Failure]
\begin{Verbatim}[commandchars=\\\{\},fontsize=\footnotesize]
def superPow(a, b):
    return pow(a, pow(10, len(b) - 1, 1140)\targetapi{[0]}, 1337)\targetapi{[0]}
\end{Verbatim}
\end{tcolorbox}
\vspace{-8pt}
\captionof{figure}{MALMEN correctly applies the S4 form, indexing both
\texttt{pow} calls with \texttt{[0]}. However, the algorithm is wrong: the exponent
is derived from \texttt{len(b)} alone, discarding the actual digit values in \texttt{b}. It loses the digit-array
decomposition required to reconstruct the exponent correctly, producing wrong outputs
and failing all tests with an \texttt{AssertionError}.}
\vspace{16pt}

\end{document}